\documentclass[preprint]{elsarticle}

\usepackage{amsfonts}
\usepackage{color}
\usepackage{amssymb}
\usepackage{amsmath}
\usepackage{graphicx,color}

\usepackage{amsthm}
\usepackage[latin9]{inputenc}
\usepackage{enumerate}
\usepackage{bm}
\usepackage{fancyhdr}

\newtheorem*{lemma*}{Lema}

\numberwithin{equation}{section}

\newcommand{\npro}{\mathcal{N}}

\newcommand{\mpro}{\mathcal{M}}

\newcommand{\opro}{\mathcal{O}}

\newcommand{\I}{\mathbb{I}}

\newcommand{\R}{\mathbb{R}}

\newcommand{\kes}{\textbf{k}}
\newcommand{\qes}{\textbf{q}}

\newcommand{\half}{\frac{1}{2}}

\newcommand{\tr}{{\rm tr}}
\newcommand{\e}{{\rm e}}

\newcommand{\li}{{\rm li}}

\begin{document}

%\begin{frontmatter}

\title{{\bf {Attractive and Repulsive Casimir Vacuum Energy with General Boundary Conditions}}}

\author{
 M. Asorey}
 \address {
Departamento de F\'{\i}sica Te\'orica, Facultad de Ciencias \\
 Universidad de Zaragoza,
  E-50009 Zaragoza, Spain }

\author{  J. M.  Mu\~noz-Casta\~neda}
 \address {Institut f\"ur Theoretische Physik, Universit\"at Leipzig, \\
Br\"uderstr. 16, D-04103 Leipzig, Germany   }

  \begin{abstract}
  
  The infrared behavior of quantum field theories confined in bounded domains
   is strongly dependent on the shape and structure of space
boundaries. The most  significant physical effect arises in the behaviour of the 
vacuum energy. The Casimir  energy can be attractive or repulsive depending on the nature of the boundary.
We calculate  the vacuum energy for a massless scalar field  confined between two 
homogeneous parallel plates  with the most general type of boundary conditions 
depending on four parameters. The analysis provides a powerful method to identify 
which boundary conditions generate attractive  or repulsive Casimir forces between the plates. 
In the interface   between  both regimes we find a   very interesting family of boundary conditions  
which do not induce any type of Casimir force. We also show that the attractive regime holds far
beyond identical boundary conditions for the two plates required by the Kenneth-Klich 
theorem and that the strongest attractive Casimir force appears for periodic boundary conditions
whereas the  strongest repulsive Casimir force  corresponds to anti-periodic boundary conditions.
Most of the analysed boundary conditions are new and some of them can be physically 
implemented with metamaterials.
\end{abstract}

%\end{frontmatter}
\maketitle

%%%%%%%%%%%%%%%%%%%%%%%%%%%%%%%%%%%%%%%%%%%%%%%%%%%%%%%%%%%
\section{Introduction.}
%%%%%%%%%%%%%%%%%%%%%%%%%%%%%%%%%%%%%%%%%%%%%%%%%%%%%%%%%%%

The  role of boundaries in quantum field theory  has been a focus of increasing activity   in different areas of physics. In general,  the presence of boundaries enhances quantum aspects of the system.  Boundary properties
have been known to play an important role in  Casimir effect \cite{casimir48}  since the early days of quantum field theory. More recently it has become a basic ingredient in the analysis of  the very first principles 
of fundamental physics: black hole quantum physics %\cite{bekenstein73,hawking74},
quantum holography, %\cite{thooft93,susskind94},
string theories and D-branes %\cite{polchinski95} 
and AdS/CFT dualities.% \cite{malda98}. 

Boundary phenomena determine the structure of the quantum vacuum and the low energy behaviour of the quantum field theories. In massless theories these effects are amplified because the existence of long distance correlations allow  boundary effects to percolate throughout the whole bulk region. In that case the vacuum energy is highly dependent on the geometry of the physical space and the physical properties of the  boundaries encoded by boundary conditions \cite{Symanzik:1981wd}-\cite{book2}.  

In this paper we focus on  the dependence of vacuum energy on boundary 
conditions in a massless field theory confined to a domain bounded by two homogeneous parallel plates.  The dependence of this energy with the distance between the plates is the basis of Casimir effect. Indeed, the variation of vacuum energy due to vacuum fluctuations  induces a force between the plates. If the plates are identical this force is attractive as demonstrates the  Kenneth-Klich theorem \cite{KK}. 
In general,  this theorem shows that due to the general principles  of quantum field theory the force induced by quantum vacuum fluctuations  between  two identical but not necessary planar bodies is always attractive. 
However, it is of enormous interest to get physical configurations where the Casimir force is repulsive instead of attractive, not
only by its relevance for technical applications to  micro-mechanical devices (MEMS),  but also because 
the existence of repulsive or null Casimir forces allows a more accurate   analysis of micro-gravity effects. There are recent conjectures about  the violation of Newton gravitational law At sub-millimeter scales (see references
\cite{hoyle}-\cite{kapner})
and to clarify the possible physical deviations at this short distances regime it is essential to disentangle  gravitational effects from Casimir force  (see references \cite{onofrio}-\cite{KMM}).  In this study the control of Casimir forces is essential  and in repulsive Casimir
regimes is easier to discriminate from gravitational effects.

All  methods used to achieve a repulsive Casimir effect are based on  plates with different
properties. In fact,  new repulsive regimes of the Casimir effect have been found between different dielectric plates \cite{capasso}, and between a metallic plate with a hole and a needle pointing to  the hole center \cite{hole}.
 In this paper we consider the most general boundary 
conditions for two plates which turn out to  depend on four parameters to analyse in great detail the transition from  attractive to repulsive Casimir regimes \cite{brevik}-\cite{ischia}.
with particular emphasis on the characterization of Casimirless boundary conditions in the interface of both
regimes \cite{orsay}.
Although in practice, only some of these boundary conditions can be physically implemented, the advances in nano-science allow to the construction of new materials (metamaterials) with very special characteristics, which may allow, in the near future, the implementation of new types of boundary conditions.

%%%%%%%%%%%%%%%%%%%%%%%%%%%%%%%%%%%%%%%%%%%%%%%%%%%%%%%%%%%
\section{Vacuum Energy of Bosonic Massless Fields in Bounded Domains}
\label{sec:1}
%%%%%%%%%%%%%%%%%%%%%%%%%%%%%%%%%%%%%%%%%%%%%%%%%%%%%%%%%%%

The infrared properties of quantum field theory are very sensitive to boundary
conditions \cite{karpacz}. In particular the physical properties of the quantum 
vacuum state and the vacuum energy exhibit  a very strong dependence on 
the type of boundary conditions.

One of the most important effects of  boundaries in field theories is the
appearance of Casimir effect. Within the global framework of boundary conditions
formulated above we can analyse with complete generality   
which boundary conditions generate attractive
or repulsive Casimir forces, i.e.  the scope of  attractive
and repulsive regimes in the Casimir effect.

Let us consider, for simplicity, a free massless complex scalar field $\psi$ confined 
in a  domain $\Omega\subset {\mathbb {R}}^D$ bounded by two parallel homogeneous plates. 
Let us assume that the parallel plates are  orthogonal to the $OX_{_D}$ direction and are placed at $x_{_D}=0$ and $x_{_D}=L$, 
respectively.
Although  physically interesting systems are  three-dimensional $(D=3)$,
for some interesting applications we also consider  two-dimensional systems $(D=2)$. 
The results can be easily  generalised for massless fermions and gauge theories.

The Hamiltonian is given by 
\begin{equation}
  {\widehat{\bf H}}={1\over 2} \int_\Omega d^{D}{\bf x} \left(|\widehat{\pi}({\bf x})|^2 - {\widehat{\psi}}^\ast ({\bf x}) {{\, \Delta\,}}\,\widehat{\psi}({\bf x})
  \right), \label{ham}
\end{equation}
with standard canonical quantization commutation rules 
\begin{equation}
[\widehat{\pi}({\bf x}), \widehat{\psi}({\bf x'})]=-i \hbar\, \delta({x-x'}),
\end{equation}
which describes  an infinite number of decoupled harmonic oscillators given by the Fourier modes of 
the operator $ -\Delta$. Unitarity requires that the Hamiltonian (\ref{ham}) has to be self-adjoint
which is the case if  all oscillating frequencies of these harmonic oscillators are real and non-negative.
This requirement can be fulfilled if and only if all eigenvalues of the Laplacian operator 
$-\Delta$  are real and nonnegative, i.e. $-\Delta$ is a  non-negative selfadjoint operator.

Because of the homogeneity of the plates the boundary conditions must be  invariant under translation
along the plates. Local boundary conditions  of physical states  $\psi$ in the  domains 
of the selfadjoint extensions of $-\Delta$ have
 been characterised in Ref. \cite{aim} in terms of  $2\times 2$ unitary matrices $U\subset U(2)$. They are given by
\begin{equation}
 \varphi - i\delta\,\dot\varphi = U(\varphi +i\delta\,\dot\varphi ),
 \label{bccc}
\end{equation}
where 
\begin{equation}
 \varphi=\begin{pmatrix}
			\varphi(L)  \\ \varphi(0)  \\ 
		\end{pmatrix},
		\quad
  \dot\varphi=\begin{pmatrix}
			\dot\varphi(L)  \\ \dot\varphi(0)  \\ 
		\end{pmatrix},
\end{equation}
are the boundary values  $\varphi= \psi|_{_{\partial\Omega}}$ of  the states  $\psi$   and  their outward normal derivatives $\dot\varphi=\partial_n\psi|_{_{\partial\Omega}}$   on the plates, and 
$\delta$ is an arbitrary characteristic length parameter.

However, non-negativity imposes a further constraint   \cite{tesis, gadella, orsay} on boundary conditions (\ref{bccc}). Indeed,
 any state $\psi$ whose boundary values $\varphi$
are eigenvalues of  the unitary operator  $U\varphi=\e^{i\alpha}\varphi$ verifies   the identity  \cite{aim}
\begin{equation*}
\langle\psi,-\Delta_U \psi\rangle= \parallel d\, \psi \parallel^2 
+\, \delta^{-1}\tan\frac{\alpha}{2}  \parallel\varphi \parallel^2,
\end{equation*}
which implies that the self-adjoint extension $-\Delta_U$ can be non-negative  for large enough volumes only if $\pi<\alpha<2\pi$.
For simplicity,   from now we shall assume  $\delta=1$

In the standard parametrization of U(2) matrices
\begin{eqnarray}
  U(\alpha,\beta,{{\bf n}})&=&\e^{i\alpha}\left(\I \cos\beta+i{{\bf n}}\cdot\bm\sigma\,\sin\beta \right);
   \label{parametrization}\quad  {\alpha\in[0,2\pi],\,\, \beta\in[-\pi/2,\pi/2]}
\end{eqnarray}
in terms of an unitary vector ${{\bf n}}\in S^2$ and Pauli matrices $\bm{\sigma}$,%($|{{\bf n}}|^2=1$)
the space of boundary conditions $\mpro_F$ which give rise to positive selfadjoint  extensions of $-\Delta$  is reduced to 
\begin{equation}
  \mpro_F\equiv\left\{U(\alpha,\,\beta,\,{{\bf n}})\in U(2)\,|\,0\leq\alpha\pm\beta\leq\pi\right\}.\label{dominiocons}
\end{equation}
since the eigenvalues of $U$ are $\e^{i(\alpha\pm\beta)}$.

The boundaries of the space $\mpro_F$ are the Cayley submanifolds ${\cal C}_\pm$\cite{aim} given by unitary operators $U$ having at least one real eigenvalue\footnote{Cayley submanifolds ${\cal C}_\pm$ have a stratified structure characterised by the multiplicities of the eigenvalues $\pm 1$} $\lambda=\pm 1$ ({$\alpha=0,\pi$}). The rich structure of this space includes very sophisticated boundary conditions which have never been considered in field theory in bounded domains. Most of the boundary conditions are non-local and some of them can be experimentally implemented coating the boundary with suitable metamaterials.
Some of them  involve topology changes \cite{bala, aim}  which motivated recent  interesting proposals  related to quantum gravity \cite{SWX,survey}.  
\begin{figure}[htbp]
  \centerline{\includegraphics[height=8cm]{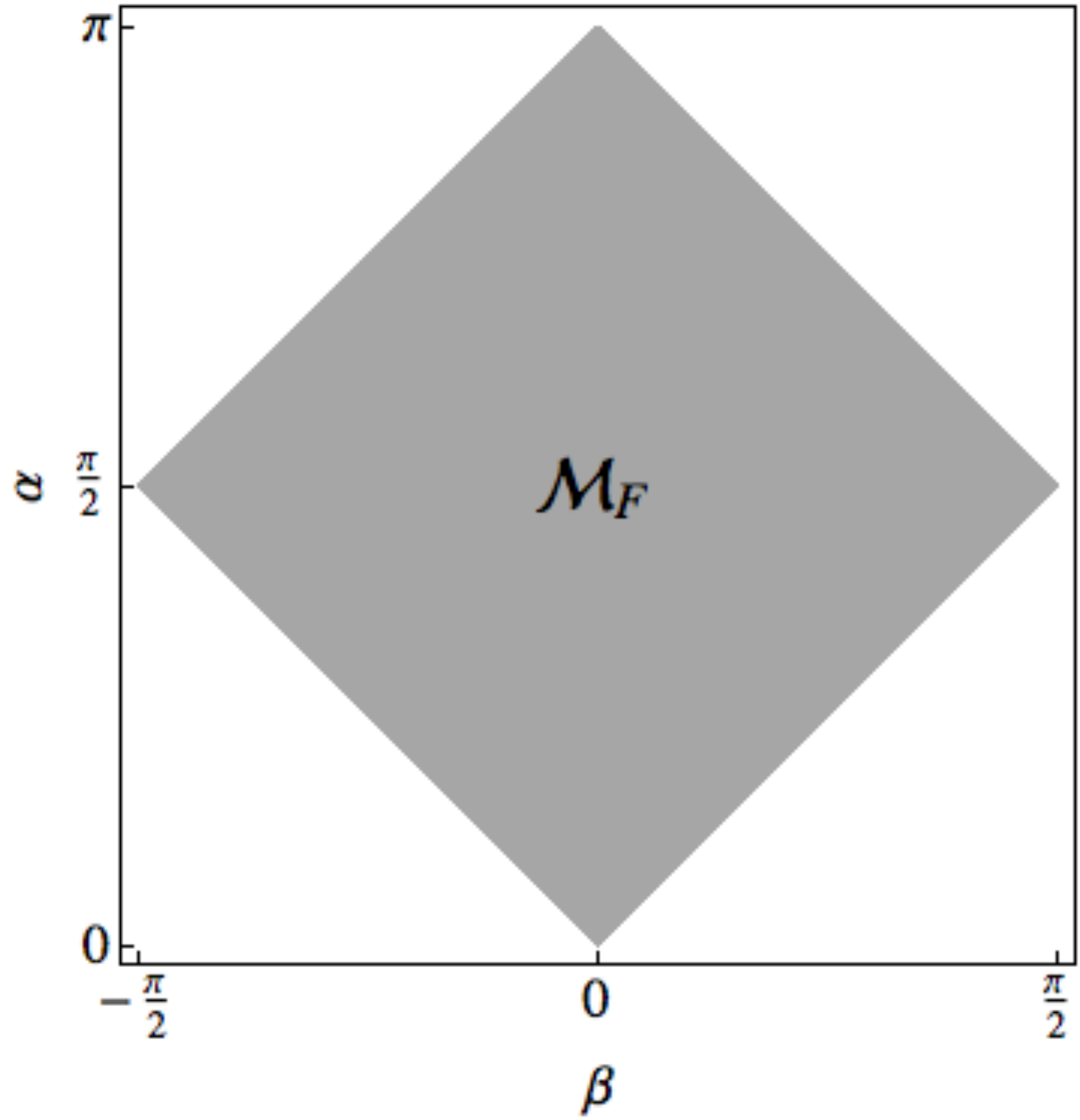}}
  \caption{\footnotesize{Rhombic slice 
  of the space of consistent boundary conditions $\mpro_F$ for a scalar field theory confined 
 between two homogeneous parallel plates in the $\alpha-\beta$ planes for fixed value of $\bf n$ and $0<\alpha\pm\beta<\pi.$} The conical structure of the full space $\mpro_F$ of boundary conditions   is clearly inferred from the displayed rhombus.} \label{rombo}
\end{figure}

%%%%%%%%%%%%%%%%%%%%%%%%%%%%%%%%%%%%%%%%%%%%%%%%%%%%%%%%%%%
%\section{Vacuum Energy.}
%\label{sec:2}
%%%%%%%%%%%%%%%%%%%%%%%%%%%%%%%%%%%%%%%%%%%%%%%%%%%%%%%%%%%

The vacuum state  of  the scalar free field theory with  boundary condition $U\in\mpro_F$
is unique and given by  
\begin{equation}
\Psi_0(\psi)= \npro \ {\rm e}^{\displaystyle - \frac12 (\psi, \sqrt{-\Delta_U}\, \psi)}
\end{equation}
in the functional Schr\"odinger
representation,  $\npro$ being a normalisation constant. 
The energy corresponding to the  Gaussian vacuum state $\Psi_0(\psi)$  is given by the sum of the eigenvalues of  $\sqrt{-\Delta_U}$, i.e.
 \begin{equation}
  E^{}_U= \tr\,   {\sqrt{-\Delta_U}}. \label{divac}
\end{equation}
Notice the absence of the $\frac12$ factor because of the complex nature of the
fields. For  scalar real  fields the Casimir energy is  $\frac12$ of the result for complex
scalars  with the restriction that only boundary conditions with $U=U^\top$ (i.e. $n_2=0$ for
parallel plates) should be considered \cite{adj1}.    In present case of conformal massless  theories the infrared properties of the theory are enhanced and the genuine Casimir effect is stronger. In this regime the dependence on the boundary conditions of the fields also becomes more significant. In fact in the case of boundary conditions with zero-modes the vacuum state becomes unbounded and ill-defined. The problem disappears if the scalar field is compactified (see refs.  \cite{aj1,aj2}). Although the contribution of the zero-modes to the boundary entropy at finite temperature is crucial \cite{aj2} they do not contribute to the vacuum Casimir energy. 
 
The sum $  \tr \sqrt{-\Delta_U}$ is ultraviolet divergent but there  are finite volume corrections to the vacuum energy density  which give rise to a finite neat Casimir effect. The divergences  can be regularized using the heat equation kernel method \cite{Kirsten:2001wz,Vassilevich:2003xt,avramidi}. Indeed, we replace the divergent expression (\ref{divac}) by 
\begin{equation}
  E^{\epsilon}_U=
   \tr\,   \sqrt{-\Delta_U} {\rm e}^{-\epsilon \Delta_U}, \label{trazareg}
\end{equation}
where $\epsilon$ is the ultraviolet regularization parameter with units of inverse energy. The field theory is defined in the physical limit $\epsilon \to 0$. Before taking the physical limit we can make an asymptotic expansion in the distance $L$ between plates to obtain the regularized expression of the vacuum energy between the plates, which behaves as 
\begin{equation}
\frac{E^{(L,\epsilon)}_U}{S}= c_0 \epsilon^{-{D/2-1/2}}L + {c_1}{} \epsilon^{-D/2} + \frac{c^{(D)}}{L^{D}} +
 \opro\left({\epsilon^\half \over L^{D+1}}\right).\label{asint}
\end{equation}
where $S$ is the (infinite) volume of the plates. In the regularized $L$ expansion (\ref{asint}) each term has a different physical meaning:
\begin{enumerate}
\item {The first term $ c_0 \epsilon^{-{D/2-1/2}}$ is the energy density of the field theory in the bulk 
    \begin{equation}
      c_0 =\frac{\epsilon^{{D/2+1/2}}}{(2 \pi)^D}  \int d^Dk\,  |k|\,  {\rm e}^{-\epsilon k^2}= \frac{
      {\Gamma(\frac{D+1}{2})}}{(4\pi)^{\frac{D}{2}} \Gamma(\frac{D}{2})}\label{ecer}
    \end{equation}}
    and does not depend on the boundary conditions but is ultraviolet divergent. 
\item {The second term $c_1 \epsilon^{-D/2}$  is the surface energy density associated to the plates. It presents a lower degree of ultraviolet divergence and depends on  boundary conditions.}
\item {The third term constitutes the first finite contribution to the vacuum energy
  \begin{equation}
    {E^{(L)}_U}=  \frac{c^{(D)}}{L^{D}} S,
  \end{equation}
  and defines the Casimir energy. The corresponding Casimir force
  \begin{equation}
    {F^{(L)}_U}= D \frac{c^{(D)}}{L^{D+1}}S
  \end{equation}
  is suppressed by the ${L^{-D-1}}$ power law and its character can be attractive or repulsive, depending on the sign of the coefficient $c^{(D)}$.}
\end{enumerate}

To determine the attractive or repulsive nature of the Casimir force we calculate the coefficient $c^{(D)}$  as a function of the consistent boundary conditions given by  unitary operators $U\in \mpro_F$.

The spectrum of $\Delta_{U}$  has a continuous component indexed by $D-1$ coordinates, whereas the compact and bounded direction orthogonal to the plates generate a discrete component in the spectrum, which will depend on the boundary condition defined by $U$, i.e.
\begin{equation*}
\lambda_{i,{\bf k}}=\kappa_i^2+\sum_{j=1}^{D-1}k_j^2,
\end{equation*}
where  ${\bf{k}}=(k_1,\cdots,k_{D-1})$
is any vector of $\R^{D-1}$ and $\kappa_i^2, i=0,1,\cdots \infty $  are the eigenvalues of the operator $$-\Delta_U^{(D)}=-\frac{d^2}{d x{_{{_D}}}\!\!\!\!^{2}}$$ 
acting  on functions defined in $[0,L]$ with 
boundary conditions (\ref{bccc}).

 Therefore the functional trace  can be written in the form
\begin{equation*}
  \tr\left(-\Delta_{U}\right)^{1/2}\e^{-\epsilon  \Delta_{U}}= \frac{S}{(2\pi)^{D-1}} \sum_{i=0}^\infty\int d^{{}^{\phantom{}D-1}}\!\!\!\!\!\!\!\!\kes\ \ \  {\rm e}^{-\epsilon (\kes^2+\kappa_i^2)} \sqrt{\kes^2+\kappa_i^2},
\end{equation*}
where $S$ is the infinite $(D-1)$-volume of the plates. Performing the change of variables $q_j=k_j/\kappa$ we obtain
\begin{equation*}
  \tr\left(-\Delta_{U}\right)^{1/2}\e^{-\epsilon \Delta_{U}}= \frac{S}{(2\pi)^{D-1}} \sum_{i=0}^\infty\kappa_i^D\int_{}d^{^{D-1}}\!\!\!\!\!\!\!\!\!\qes\ \    {\rm e}^{-\epsilon  \kappa_i^2 (\qes^2+1)}\sqrt{\qes^2+1},
\end{equation*}
and using generalized spherical coordinates, once the angular variables are integrated out, we obtain 
\begin{equation*}
  E^{(D)}_U {(\epsilon)}=\frac{S\, \Omega_{D-2}}{(2\pi)^{D-1}}\sum_{i=0}^\infty \kappa_i^D \int_0^\infty dq\,\e^{-\epsilon\kappa_i^2 (q^2+1)}q^{D-2} \sqrt{q^2+1} ,
\end{equation*}
where  
\begin{equation*}
  \Omega_{D-2}=2\frac{\pi^{\frac{D-1}{2}}}{\Gamma\left(\frac{D-1}{2}\right)},
\end{equation*}
 is the area of the $(D-2)$-sphere.
The  $q$-integral    can be  expressed in terms of
the confluent hypergeometric function $U(a,\,b,\,z)$ as
\begin{equation}
\int_0^\infty dq\ \e^{-\epsilon \kappa^2 (q^2+1)}\, q^{D-2}\sqrt{q^2+1}= \half{\Gamma\left(\frac{D-1}{2}\right)}\, U\left({D-1\over 2},\,\frac{D}{2}+1,\,\epsilon\kappa^2\right)\ \e^{-\epsilon  \kappa^2},
\end{equation}
for any values of $D>0$ and $\epsilon>0$. Therefore the regularised vacuum energy per unit volume of the plates can be written as
\begin{equation}
  \frac{E^{(D)}_U {(\epsilon)}}{S} =  \frac{1}{2^{D-1}\pi^{\frac{D-1}{2}}}\sum_{i=0}^\infty \kappa_i^D\,\e^{-\epsilon  \kappa_i^2} U\left({D-1\over 2},\,\frac{D}{2}+1,\,\epsilon\kappa_i^2\right).\label{iexac}
\end{equation}

To perform the sum we have to calculate the eigenvalues $\kappa^2_i$. 
They can be found by  imposing the boundary conditions (\ref{bccc}) on the wave functions
of the domain of  the selfadjoint extension $ -\Delta_U^{(D)}$ 
\begin{eqnarray}
  \psi_k(x)&=&C_1\e^{-ikx}+C_2\e^{ikx}\label{fcs},
\end{eqnarray}
which give rise to a linear homogeneous equation
for the coefficients $C_1$ and $C_2$,
\begin{equation}
  (M-UN)\left(\begin{array}{c} C_1+C_2 \\ C_1-C_2 \\ \end{array}\right)=0\label{sl}
\end{equation}
where $M$ and  $N$ are the  $2\times2$ complex matrices
\begin{equation}
  M=\left(\begin{array}{cc}
  1 & -k \\ \cos{kL}+ik\sin{kL} & k\cos{kL}+i\sin{kL} \\
  \end{array}\right)\label{A}
\end{equation}
\begin{equation}
  N=\left(\begin{array}{cc}
  1 & k \\ \cos{kL}-ik\sin{kL} & -k\cos{kL}+i\sin{kL} \\
  \end{array}\right)\label{B}.
\end{equation}
Matrices $M$ and $N$ are the linear maps from the scattering data into the boundary values. 
These maps will become infinite dimensional in  higher dimensional space-times.
The linear system  (\ref{sl}) has non-trivial solutions if and only if $\det(M-UN)=0$. Thus,  
the eigenvalues  $\kappa_i^2$ of $ -\Delta_U^{(D)}$  are given by the zeros of the  spectral function (see reference \cite{brevik}) 
\begin{equation*}
\begin{array}{lll}
\displaystyle  h_U(k)=\det(M-UN)\!\!\!\!&=&\!\! \!\! 4 k \det U\cos kL- 2i (1+k^2) \det U\sin kL + 4
 k (U_{21}+U_{12})  \phantom{}
\cr
\displaystyle &&\!\!\!\! -2 i (1+k^2) \sin kL -4 k \cos kL  + 2 i (1-k^2) \tr U \sin kL.\phantom{\Biggr|}
\end{array}
\end{equation*}
 These zeros  provide the eigenvalues of
$-\Delta_U^{(D)}$ with one exception: the zero modes $k=0$. In this case the maps $M$ and $N$ have to be modified because
the plane wave parametrisation of scattering becomes degenerate and does not account for all possible zero-mode eigenfunctions. However the information about the zero-modes of $ -\Delta_U^{(D)}$ for any $U\in{\cal M}_F$ is encoded in $h_U(k)$ \cite{mckk}. Using parametrisation (\ref{parametrization}) we can write the spectral function $h_U(k)$ as
\begin{eqnarray}
h_U(k)&=&\nonumber 2i e^{i\alpha}\left[ \sin(kL)\left((k^2-1)\cos(\beta)+(k^2+1)\cos(\alpha)\right)\right.\\
&&\!\!\!\!\!\!-\left.2k\sin(\alpha)\cos(kL)-2k n_1\sin(\beta)\right].\label{hspec-param}
\end{eqnarray}
The apparent lack of dimensional homogeneity of the $k$ powers is due to the fact that we have 
chosen $\delta=1$ in the boundary conditions (\ref{bccc}). In general  they should be  considered as powers
of the dimensionless variable $ k\, \delta$.

The spectral function is not only dependent on
the algebraic invariants of the boundary unitary matrix $\det\, U$ and $\tr\, U$ but also on the entries
$U_{21}$ and $U_{12}$, which implies that the spectrum of
the quantum theory will be different for $U$ matrices with the same eigenvalues even when they are
equivalent as matrices. Of note is that all the zeros of the spectral function $h_U$ lay on the
positive real line of the complex  $k$-plane because the consistency conditions ensure the non-negativity of the selfadjoint extension $ -\Delta_U^{(D)}$ when $U\in{\cal M}_F$.

The sum over the eigenvalues of the operator $({-\Delta_U^{(D)}})^{\half}$ is equivalent to the sum over the  zeros of the spectral function $h_U(k)$. Null eigenvalues which are incorrectly described by $h_U$ do not contribute in both cases. Since $h_U(k)$ is holomorphic in $k$, using the  argument principle,  the summation over zeros can be rewritten in terms of an contour integral enclosing all the zeros of $h_U$ (see Refs.\cite{watson,somm,vkns, Kirsten:2001wz,Vassilevich:2003xt} for more details).   Indeed because of the consistency conditions all  zeros of $h_U$ are contained in $\R^+$ 
and by using Cauchy's residues theorem  
\begin{equation}
  \frac{E^{(D)}_U {(\epsilon)}}{S}=\sum_{n=0}^\infty k_n \e^{-\epsilon \kappa_n^2}  {\textstyle U\left({D-1\over 2},\,\frac{D}{2}+1,\,\epsilon\kappa_i^2\right)}
={1\over 2\pi i}\oint dk\ k\ {\rm e}^{-\epsilon k^2}  {\textstyle U\left({D-1\over 2},\,\frac{D}{2}+1,\,\epsilon k^2\right)} \frac{d}{dk}\log(h_U(k))\label{intc},
\end{equation}
where the integration contour encloses a thin strip around all the positive real axis %(see Figure \ref{contorno}) 
which includes all the zeros of $h_U$. The zeros of the spectral function generate simple or double poles (depending on eigenvalue degeneracies) of the logarithmic derivative $\frac{d}{dk}\log h_U(k)$. There are many similar integral formulas in the Casimir literature \cite{scatt}-\cite{kw}(see \cite{milton} for more complete list of references, and \cite{Kirsten:2001wz} for a review of spectral techniques in quantum field theory), but
most of them do not apply to the very general type of boundary conditions that we are considering.
%\begin{figure}[htbp]
%\centerline{\includegraphics[height=2.5cm]{contorno.pdf}}
%\caption{\footnotesize{Contour  enclosing the zeros of $h_U(k)$ which allows the calculation of the Casimir energy as a contour integral. }}\label{contorno}
%\end{figure}

In the limit $\epsilon\to 0$ the expression (\ref{intc}) diverges as indicated by the asymptotic expansion (\ref{asint}). To extract the finite part of this expression, that contains the Casimir energy, we have to subtract not only the leading divergence of the vacuum  energy induced from fluctuations of the fields in the bulk but also remove the subleading divergent contribution associated with the self-energy of the boundaries. The later can be achieved by  subtracting  the vacuum energy of an identical system with the same boundary condition {defined over} a fixed reference size $L_0<L$. After both subtractions 
a finite value for  the Casimir energy is obtained
\begin{equation}
  \frac{{c}_U^{(D)}}{L^{D}}=\frac{L_0^D}{L^D-L_0^D}\lim_{\epsilon\to 0}\left(c^{(D)}_0(\epsilon)(L-L_0)- \frac1S(E^{(L)}_U(\epsilon)-E^{(L_0)}_U(\epsilon))\right),\label{asincas}
\end{equation}

Thus, the Casimir energy  is given by the finite part of (\ref{intc}) which can be obtained by first removing the  $\epsilon^{-\frac{D+1}2}$ and $\epsilon^{-\frac{D}2}$ divergent terms of $E^{(D)}_U(\epsilon)$ and then taking the physical limit $\epsilon\rightarrow 0$. The behaviour of the finite contribution of (\ref{intc}) is strongly dependent on the parity, even or odd, of the spacial dimension $D$ because of the different asymptotic behaviors of the confluent hypergeometric function $U(a,\,b,\,z)$.

In the odd case $D=2n+1$ the leading  behavior of the non-divergent part $ U_+\left({D-1\over 2},\,\frac{D}{2}+1,\,\epsilon\kappa^2\right)$ of the  $\epsilon\kappa$-expansion of  $ U\left({D-1\over 2},\,\frac{D}{2}+1,\,\epsilon\kappa^2\right)$   is a constant term \cite{Abramowitz,nistdl}
\begin{equation}
U_+\left({n},\,n+\frac{3}{2},\,\epsilon\kappa^2\right)= -\frac{\Gamma\left(-n-\half \right)}{2 \sqrt{\pi}}+{\cal O}(\epsilon \kappa^2),
\label{asyodd}
\end{equation}
whereas in the  even case $D=2n$ there is a leading logarithmically divergent term \cite{Abramowitz,nistdl}
\begin{equation}
U_+\left(n-{1\over 2},\,{n}{}+1,\,\epsilon\kappa^2\right)= \frac{ (-1)^{n}}{2\sqrt{\pi}\,\Gamma(n+1)}\left(\psi\left(n-1/2\right)-\psi(n+1)+\gamma+  \log(\epsilon \kappa^2)\right)+{\cal O}(\epsilon \kappa^2),
\label{asyeven}
\end{equation}
 $\psi(s)=\frac{\Gamma'(s)}{\Gamma(s)}$ being the digamma function and  $\gamma$  the Euler constant.

One important fact is that  there are no cancellations in the expression  (\ref{iexac}) between 
 sub-leading asymptotic terms ${\cal O}(\epsilon \kappa_i^2)$ of the 
confluent hypergeometric function $
U\left(n-{1\over 2},\,{n}{}+1,\,\epsilon\kappa_i^2\right)$ 
and divergent contributions coming from the remaining sum over
 the eigenvalues $\kappa_i$; nor between  divergent terms of  $
U\left(n-{1\over 2},\,{n}{}+1,\,\epsilon\kappa_i^2\right)$ and subleading   ${\cal O}(\epsilon)$
terms of the sums over $\kappa_i$.

 In summary, the Casimir energy arises only from the product
of the finite  ${\cal O}(1)$ terms of  (\ref{asyodd}) and (\ref{asyeven}) which correpond to two different types of
  behaviours.

\subsection*{\bf a) Odd dimensional  $D=2n+1$ spaces.} 

In this case the Casimir  energy  (\ref{asincas}) can be obtained from  formula  (\ref{intc}) keeping only  the leading 
asymptotic contributions  (\ref{asyodd})  of the  confluent hypergeometric function  ${\textstyle U\left({D-1\over 2},\,\frac{D}{2}+1,\,\epsilon k^2\right)}$.
%U\left(n-{1\over 2},\,{n}{}+1,\,\epsilon\kappa_i^2\right)$ (\ref{asyodd}) 
%%is of the form
%%\begin{equation*}
%%\sum_{i=0}^\infty \kappa_i^D \e^{-\epsilon \kappa_i^2}
%%\end{equation*}
%%
%%\begin{equation}
%%  \displaystyle
%%  \frac{c^{(2n+1)}_U}{L^{{2n+1}}}\!\!=\!\!\displaystyle  \frac{(-1)\Gamma\left(-\frac{2n+1}{2}\right)\, L_0^{2n+1}}{(4\pi)^{\frac{2n+1}{2}} (L^{2n+1}\!-\!L_0^{2n+1})} \lim_{\epsilon\to 0 }
%%  \sum_{i=0}^\infty \kappa_i^D \e^{-\epsilon \kappa_i^2}
%%   \frac{1}{ 2\pi i }\oint dk\, k^{2n+1}\, {\rm e}^{-\epsilon k^2} \left[ {(L_0\!- \!L) \frac{ k\!-\! k^\ast}{|k\!-\!k^\ast|}}
%%  \right. 
%%  \displaystyle
%% \! -\!\left.
%%   \frac{d}{dk} \log\left(\frac{h_U^{(L)}(k)} {h_U^{(L_0)}(k)}\right)
%%   \right].
%%\end{equation}
The result is given by the following  contour integral of  the spectral function
%
%
%the sum contribution is of the form
%\begin{equation*}
%\sum_{i=0}^\infty \kappa_i^D \e^{-\epsilon \kappa_i^2}
%\end{equation*}
% and can be estimated by the method already used in  the $1+1$ dimensional case. Indeed, the sum over the zeros can be written as a contour integral
%\begin{equation}
% \sum_{i=0}^\infty    \kappa_i^D \e^{-\epsilon \kappa_i^2}={ 1\over 2\pi i}\oint  dk \,k^D  \e^{-\epsilon k^2} \frac{d}{dk}\log(h_U(k))\label{sumcero}
%\end{equation}
%where the integration contour encloses all zeros of the spectral function $h_U$  on the real positive axis (Figure \ref{contorno}). The integral diverges in the physical limit $\epsilon\to 0$. To calculate its finite contribution one has to first  subtract, as in the $1+1$-dimensional case,  the same contribution for a reference size $L_0$ to remove the UV divergent surface energy terms,  and later to subtract the leading divergence terms coming from the bulk energy. The Casimir energy  is then recovered as  the $\epsilon\to 0$ limit
%\begin{equation}
%  \frac{{c}_U^{(D)}}{L^{D}}=\frac{L_0^D}{L^D-L_0^D}\lim_{\epsilon\to 0}\left(c^{(D)}_0(\epsilon)(L-L_0)- \frac1S(E^{(L)}_U(\epsilon)-E^{(L_0)}_U(\epsilon))\right),\label{asincas}
%\end{equation}
%which can be rewritten as 
\begin{equation}
  \displaystyle
  \frac{c^{(2n+1)}_U}{L^{{2n+1}}}\!\!=\!\!\displaystyle  \frac{(-1)\Gamma\left(-\frac{2n+1}{2}\right)\, L_0^{2n+1}}{(4\pi)^{\frac{2n+1}{2}} (L^{2n+1}\!-\!L_0^{2n+1})} \lim_{\epsilon\to 0 } \frac{1}{ 2\pi i }\oint dk\, k^{2n+1}\, {\rm e}^{-\epsilon k^2} \left[ {(L_0\!- \!L) \frac{ k\!-\! k^\ast}{|k\!-\!k^\ast|}}
  \right. 
  \displaystyle
 \! -\!\left.
   \frac{d}{dk} \log\left(\frac{h_U^{(L)}(k)} {h_U^{(L_0)}(k)}\right)
   \right].
\end{equation}

Since this expression is finite and convergent in the $\epsilon\to 0$ limit, we can drop the regulating exponential heat kernel  factor ${\rm e}^{-\epsilon k^2}$. In this case, because of the holomorphic properties of the integrand, the integration can also be extended to the contour given by an infinite semi-circle limited in its left hand side by the imaginary axis. %(see Figure \ref{circuito})
 As long as the integration over the semicircle is zero, the integration is reduced to the imaginary axis of the complex $k$-plane, and taking into account the parity invariance of the integrand the integration range can be reduced to the positive imaginary axis. The final integral expression for the Casimir energy   is
\begin{equation}
  \frac{c^{(2n+1)}_U}{L^{2n+1}}=\displaystyle{
 \frac{4(-1)^{n} \Gamma\left(-\frac{2n+1}{2}\right)\, L_0^{2n+1}}{(4\pi)^{\frac{2n+3}{2}}  (L^{2n+1}- L_0^{2n+1})} \int_0^\infty dk\ k^{2n+1}}
\left[L^{}-L_0^{}-\frac{d}{dk}\log\left(\frac{h_U^{(L)}(ik)} {h_U^{(L_0)}(ik)}\right)\right].\label{casimir3}
\end{equation}
\par
\subsection*{\bf b) Even dimensional  $D=2n$ spaces.} 
The even case presents some further interesting peculiarities. Following the analysis of the odd case, but taking into account the
different asymptotic behaviour  (\ref{asyeven}) of the    confluent hypergeometric function $U_+\left(n-{1\over 2},\,{n}{}+1,\,\epsilon\kappa^2\right)$,
%The remaining sum over the transverse discrete 
%modes involves a logarithmic term coming from the asymptotic expansion of the %confluent 
%hypergeometric function
%$U\left(n-{1\over 2},\,{n}{}+1,\,\epsilon\kappa_i^2\right)$.  for the sum
%\begin{equation*}
% \sum_{i=0}^\infty \kappa_i^{2n} \left(\psi\left(n-1/2\right)-\psi(n+1)+\gamma+  \log(\epsilon \kappa_i^2)\right) \e^{-\epsilon \kappa_i^2}
%\end{equation*}
we obtain the Casimir energy in terms of the contour integral  
\begin{eqnarray*}
  \displaystyle
  \frac{c^{(2n)}_U}{L^{2n}}= \displaystyle \frac{ (-1)^{n}(4{\pi})^{-n}\, L_0^{2n}} {  \Gamma(n+1) (L^{2n}-L_0^{2n})}\!\!\!\!\!\! &\!\!\!\!\!\!&\!\!\! \lim_{\epsilon\to 0} \displaystyle {1\over  2\pi i }\oint dk\, k^{2n}\!\!  \left[( L_0-L)\frac{ k}{|k|}- \frac{d}{dk}\log\frac{h_U^{(L)}(k)} {h_U^{(L_0)}(k)}\right]   \nonumber  \\
&&\!\!\!\!\!\!\!\!\!\!\!\!\!\!\!\!\!\!\!\!\!\!\!\! \left(\psi\left(n-\frac12\right)-\psi(n+1)+\gamma+  \log(\epsilon k^2)\right) {\rm e}^{-\epsilon k^2}. 
\end{eqnarray*}
 The exponential factor of the heat kernel can again be dropped and the integration reduced to an integral over the imaginary axis. But in this case the terms proportional   to $\psi\left(n-\frac12\right)-\psi(n+1)+\gamma$ are parity odd and the integral over the positive
and negative imaginary axes cancel each other. Only the terms proportional to the logarithm term $ \log(\epsilon \kappa_i^2)$ provide a non-vanishing contribution.  Due to the existence of a branch cut  that we fix along the real positive axis  the  contribution of  positive imaginary axis picks up a factor $i\pi/2$ whereas in the negative imaginary axis this factor is $-i\pi/2$ . 
 The total contribution of the integral reduces to a very compact finite formula
\begin{equation}
  \frac{c^{(2n)}_U}{L^{2n}}={\displaystyle - \frac{{ }(4{\pi})^{-n}\, L_0^{2n} }{  \Gamma(n+1) (L^{2n}\!-\!L_0^{2n})}}
  \int_0^\infty dk\, k^{2n} 
  \left[L-L_0-\frac{d}{dk}\log\left(\frac{h_U^{(L)}(ik)} {h_U^{(L_0)}(ik)}\right)\right]\label{casimir2n}
\end{equation}
  for the calculation of the Casimir energy.

Expressions (\ref{casimir3}) and (\ref{casimir2n}) allow the calculation of  Casimir energy for arbitrary consistent boundary conditions in the parallel plates configuration in any spatial dimension.  We restrict our analysis to the two most interesting cases from physical point of view:  massless scalar theories in $2+1$ and $3+1$ dimensional space-times (the case 1+1 has been already analysed from this
viewpoint in Refs. \cite{brevik,tesis}). What is interesting about these two cases is that many of the new boundary conditions of 
the parallel plates can be implemented in the laboratory and  the theoretical results falsified.

\section{Casimir energy in three dimensions}
\label{sec:8}
%%%%%%%%%%%%%%%%%%%%%%%%%%%%%%%%%%%%%%%%%%%%%%%%%%%%%%%%%%%

In the  most realistic three-dimensional ($D=3$) case  the  Casimir energy is given by the integral
\begin{equation}
  \frac{{E}^{(3)}_U}{S}=\frac{{c}^{(3)}_U}{L^3}=\frac{-L_0^3}{6\pi^2 (L^3-L_0^3)}\int_0^\infty dk\,k^3\left[L-L_0-\frac{d}{dk} \log\left(\frac{h_U^{(L)}(ik)}{h_U^{(L_0)}(ik)} \right)\right],\label{cas3}
\end{equation}
 for any  type of  consistent boundary conditions (i.e. $U\in\mpro_F$). 
 
 In many cases the calculation of the Casimir energy using the spectral formula (\ref{cas3}) can be achieved analytically but in general it requires the use of  numerical simulations. Let us first  compare  the analytic results obtained by the spectral function  method and  the results obtained by other methods (e.g. zeta function regularisation \cite{Kirsten:2001wz,avramidi,elizb,Blau:1988kv}).
 
% In some cases it is possible to obtain explicit analytical expressions for the Casimir energy. 
%In these cases we can also compare with the results obtained by other methods like the zeta function regularization method. These cases in which analytical results can be obtained are shown below.
\paragraph{{\rm i)} Periodic boundary conditions}They correspond to a folding of the  space into a cylinder,
 $\psi(0)=\psi(L), \psi'(0)=\psi'(L)$,
  and are  described by the unitary operator
\begin{equation}
  U_{ p}=\sigma_1=
  \left(\begin{array}{cc}
  0 & 1 \\ 1 & 0 \\
  \end{array}\right)
.\label{periodicas}
\end{equation}
The associated  spectral function,
\begin{equation}
h^{(L)}_p(k)=4k(\cos{kL}-1),
\end{equation}
reduces  the integrand in expression (\ref{cas3}) to
\begin{equation*}
L-  L_0-{}\frac{d}{dk}\left(\log\frac{h^{(L)}_p(ik)}{h^{(L_0)}_p(ik)}\right)= {L-L_0+L_0\coth{kL_0\over 2}-L\coth\frac{kL}{2}},
  \end{equation*}
and
the Casimir energy  is, thus, given by 
\begin{equation*}
  \frac{{E}^{(3)}_p}{S}=\frac{{c}^{(3)}_p}{L^3}={- L_0^3\over 6\pi^2  (L^3-L_0^3)}\int_0^\infty \,dk\, k^3\left[L-L_0+L_0\coth\left({kL_0\over 2}\right)-L\coth\left({kL\over 2}\right)\right]
\end{equation*}
which, can be  analytically integrated out, giving rise to a  negative Casimir energy,  that agrees with results obtained by other standard methods (e.g. zeta function regularization \cite{elizb}).
 \begin{equation}
  \frac{{E}^{(3)}_p}{S}=-\frac{\pi^2}{45 L^3}.
\end{equation}
As it is well known the Casimir effect  in the periodic case introduces an attractive force which tends to shrink the cylinder.
\par
\paragraph{{\rm ii)} Dirichlet boundary condition}   In this case 
$\psi(0)=\psi(L)=0$, the unitary operator 
 is $U_d=-\I$, and the associated spectral function is
\begin{equation}
h^{(L)}_d(k)=4i\sin{kL}.\label{hdirichlet}
\end{equation}
The Casimir energy is given by  the well-known Casimir result 
\begin{equation*}
  \frac{{ E}^{(3)}_d}{S}={-  L_0^3\over 6\pi^2  (L^3-L_0^3)}\int_0^\infty \,dk\, k^3\left[L-L_0+L_0\coth\left(kL_0\right)-L\coth\left(kL\right)\right]
  =-\frac{\pi^2}{720 L^3},
\end{equation*}
which again is negative and $\frac1{16}$ times smaller than the periodic case.
\paragraph{{\rm iii)} Neumann boundary condition} 
In this case, $\psi'(0)=\psi'(L)=0$, the associated unitary operator  is $U_n=\I$, and its spectral function
\begin{equation}
h^{(L)}_n(k)=4ik^2\sin{kL}.
\end{equation}
Although the spectral function $h_n(k)$ is different from $h^{(L)}_d(k)$, the result is the same as for Dirichlet boundary conditions

%
%The Casimir energy is 
%the same  as in the Dirichlet case
\begin{equation}
  \frac{ { E}^{(3)}_n}{S}=-\frac{\pi^2}{720 L^3}.
\end{equation}
In both cases, Dirichlet and Neumann, the character of the Casimir force between plates is attracive, and $\frac1{16}$ times smaller  than in the periodic case.
\paragraph{{\rm iv)} Anti-periodic boundary conditions}
They correspond  to
 $\psi(0)=-\psi(L), \psi'(0)=-\psi'(L)$,
  and are  described by the unitary operator
\begin{equation}
  U_{ap}=-\sigma_1=\left(\begin{array}{cc}
  0 & -1 \\ -1 & 0 \\
  \end{array}\right)
  \end{equation}
The associated  spectral function,
\begin{equation}
h^{(L)}_{ap}(k)=4k(\cos{kL}+1),
\end{equation}
reduces  the integrand in expression (\ref{cas3}) to
\begin{equation*}
L-L_0+L_0\tanh\left({kL_0\over 2}\right)-L\tanh\left({kL\over 2}\right).
\end{equation*}
 In the case of anti-periodic boundary conditions the Casimir energy 
\begin{equation*}
   \frac{{ E}^{(3)}_{ap}}{S}={-  L_0^3\over 6\pi^2(L^3-L_0^3)}\int_0^\infty dk\, k^3\left[L-L_0+L_0\tanh\left({kL_0\over 2}\right)-L\tanh\left({kL\over 2}\right)\right]=\frac{7 \pi^2}{360 L^3}
\end{equation*}
 is positive, which corresponds to a repulsive Casimir force between plates.
\paragraph{{\rm v)} Zaremba boundary condition} 
There are two special boundary conditions which are Neumann
at one boundary and Dirichlet at the other, or viceversa. The unitary matrices are 
$$U_Z=\pm\sigma_3=\pm \left(\begin{array}{cc}
1&  0 \\ 0  &-1
\end{array}
\right), 
$$
their spectral function 
\begin{equation}
  h_Z(k)=-8 k\cos{kL}\label{zaremba}
\end{equation}
and the corresponding Casimir energy 
%This is another case where the Casimir energy is positive, as in $1+1$ and $2+1$ dimensions,
\begin{equation*}
   \frac{{ E}^{(3)}_{Z}}{S}={-  L_0^3\over 6\pi^2  (L^3-L_0^3)}\int_0^\infty dk\, k^3\left[L-L_0+ L_0\tanh\left({kL}\right)-L\tanh\left({kL}\right)\right]=\frac{7 \pi^2}{5760 L^3},
\end{equation*}
is positive which corresponds to a  repulsive Casimir force between the parallel plates.
\par
\paragraph{{\rm vi)} Quasi-periodic boundary conditions} 
This is a one-parameter family of boundary conditions defined by
$$\psi(L)=\tan\frac{\alpha}{2}\ \psi(0),\ \psi'(L)=\cot\frac{\alpha}{2}\ \psi'(0),$$
with unitary operator 
\begin{equation}
  U_{qp}=\cos{\alpha}\sigma_3+\sin{\alpha}\sigma_1,\qquad \alpha\in[-\pi/2,\pi/2]. \label{cuasiu}
\end{equation}
The associated spectral function is
\begin{equation}
  h^{(L)}_{qp}(k)=4k(\cos{kL}-\sin{\alpha})\label{hcuasi}.
\end{equation}
and the Casimir energy
 is given by the following integral expression
\begin{equation}
 \frac{ {E}^{(3)}_{qp}}{S}={- L_0^3\over 6\pi^2  (L^3-L_0^3)}
  \int_0^\infty dk\, k^3\left[L-L_0+ \frac{L_0\sinh(kL_0)}{\sin(\alpha)-\cosh(kL_0)} -\frac{L\sinh(kL)}{\sin(\alpha)-\cosh(kL)}\right],
\end{equation}
which  gives 
\begin{equation}
  {c}^{(3)}_{qp}=-{1\over \pi^2}\left(\li_4(-i\e^{i\alpha})+ \li_4(i\e^{-i\alpha})\right),
\end{equation}
where
\begin{equation*}
  \li_n(z)\equiv\frac{(-1)^{n-1}}{(n-2)!}\int_0^1\frac{dt}{t}\log(1-zt)\log^{n-2}(t)= \sum_{j=1}^\infty\frac{z^j}{j^n}
\end{equation*}
denotes the integral logarithm function.
%\begin{figure}[htbp]
%  \centerline{\includegraphics[height=5.5cm]{New3Df-qp.pdf}}
%  \caption{\footnotesize{$\alpha$-dependence of the $c^{(3)}_{qp}$ coefficient  of  Casimir energy for the quasi-periodic boundary conditions for $\alpha\in[-\pi/2,\,\pi/2]$.}} \label{cuasi3d}
%\end{figure}
The combination of integral logarithms $\li_4(-i\e^{i\alpha})+\li_4(i\e^{-i \alpha})$ can be reduced to a fourth order polynomial in $\alpha$ for $\alpha\in[-\pi/2,\,\pi/2]$), thus the Casimir energy coefficient ${c}^{(3)}_{qp}$ is
given by 
\begin{equation}
  {c}^{(3)}_{qp}=\frac{7\pi^2}{5760}-\frac{\pi\alpha}{16}-\frac{\alpha^2}{48} +\frac{\alpha^3}{12\pi}+\frac{\alpha^4}{24\pi^2};\quad\alpha\in[-\pi/2,\,\pi/2]. \label{cuasi3dpol}
\end{equation}
The behaviour of the  coefficient $c^{(3)}_{qp}$ of the Casimir energy as a function of $\alpha$ in the interval $[-\pi/2,\,\pi/2]$
%is displayed in Figure \ref{cuasi3d} where we can see 
shows that there is a value $\alpha_0^{(qp)}$ of $\alpha$ where the Casimir energy and the Casimir force between plates vanish. This special value 
\begin{equation}
  \alpha_0^{(qp)}=\pi\left(-{1\over2}+\sqrt{1-2\sqrt{{2\over 15}}}\right)
\end{equation}
 corresponds to the splitting point between attractive and repulsive regimes. 
For $-\pi/2\leq\alpha<\alpha_0^{(qp)}$ the Casimir energy is positive and hence the Casimir force between plates has a repulsive character, and when $\pi/2\geq\alpha>\alpha_0^{(qp)}$, the Casimir force between plates becomes attractive, corresponding to a negative Casimir energy.
\par
\paragraph{{\rm vii)} Pseudo-periodic boundary conditions}
Pseudo-periodic boundary conditions are a family of one-parameter family of boundary conditions that generalise periodic and anti-periodic conditions, i.e. 
$$\psi(L)={\rm e}^{-i \alpha}\ \psi(0),\ \psi'(L)={\rm e}^{-i \alpha}\ \psi'(0).$$ 
The unitary matrices defining the boundary conditions of this family  are
\begin{equation}
  U_{pp}=\cos\alpha\, \sigma_1-\sin\alpha\, \sigma_2 =\left(\begin{array}{cc}
  0 & \e^{i\alpha} \\ \e^{-i\alpha} & 0 \\
\end{array}\right);\qquad\alpha\in[-\pi,\pi]\label{pseudou}
\end{equation}
 and their spectral functions read 
\begin{equation}
  h^{}_{pp}=4k(\cos{kL}-\cos{\alpha})\label{hpseudo}.
\end{equation}
The calculation of Casimir energy 
\begin{equation*}
 \frac{ {E}^{(3)}_{pp}}{S}={- L_0^3\over 6\pi^2  (L^3-L_0^3)}\int_0^\infty\left[L-L_0- \frac{ L \sinh(kL)}{\cosh(kL)-\cos(\alpha)}- \frac{L_0\sinh(kL_0)}{\cosh(kL_0)-\cos(\alpha)}\right] k\,dk
\end{equation*}
can be reduced to that of the   quasi-periodic boundary conditions case  by replacing $\alpha_{qp}$ with $\alpha_{pp}+\pi/2$, i.e. 
\begin{equation}
 {c}^{(3)}_{pp}(\alpha)= -\frac{\pi ^2}{45}+\frac{\alpha^2}{6}-\frac{|\alpha|^3}{6 \pi }+\frac{\alpha^4}{24 \pi ^2}; \quad
 \alpha\in[-\pi,\pi].
  \label{pseudo3dpol}
\end{equation}
%(see Figure \ref{pseudo3d}).
%\begin{figure}[htbp]
%  \centerline{\includegraphics[height=5.5cm]{New3Df-pp.pdf}}
%  \caption{\footnotesize{$\alpha$-dependence  of the ${c}^{(3)}_{pp}$ coefficient  of the Casimir energy for pseudo-periodic boundary conditions for $\alpha\in[-\pi,\,\pi]$.}} \label{pseudo3d}
%\end{figure}
There are two values of $\alpha$ where the Casimir energy  and  Casimir force between plates vanish
\begin{equation}
 \alpha_{0\pm}^{(pp)}=\mp\pi\left(1-\sqrt{1- 2\sqrt{{2\over 15}}}\right).
\end{equation}
For $\alpha_{0-}^{(pp)}< \alpha<\alpha_{0+}^{(pp)}$ the Casimir energy is negative, which leads to an attractive Casimir force between plates. However,  for  $-\pi<\alpha<\alpha_{0-}^{(pp)}$ or $\alpha_{0+}^{(pp)}<\alpha<\pi$,
the Casimir energy is positive, and the force between plates is repulsive.

\par
\paragraph{{\rm viii)} Robin boundary conditions}  
The one-parameter family of Robin boundary conditions 
$$\psi'(0)=\tan\frac{\alpha}{2}\ \psi(0),\ \psi'(L)=\tan\frac{\alpha}{2}\ \psi(L),$$
is characterised by the family of unitary matrices $U_r=\e^{i\alpha}\I$, with $\alpha\in[0,\pi]$ and spectral function 
\begin{equation}
h^{(0)}_{U_r}(k)=\nonumber{2i\,\e^{i\alpha}\left(-2k\sin\alpha\cos{kL}\right.}
+\left.\left( k^2-1+(k^2+1)\cos\alpha \right)\sin{kL}\right).\label{hrobin}
\end{equation}

In this case it is not possible to find an analytical expression for the Casimir energy ${c}^{(3)}_r$, thus one has to proceed numerically from expression (\ref{cas3}). 
\begin{figure}[htbp]
  \centerline{\includegraphics[height=5.5cm]{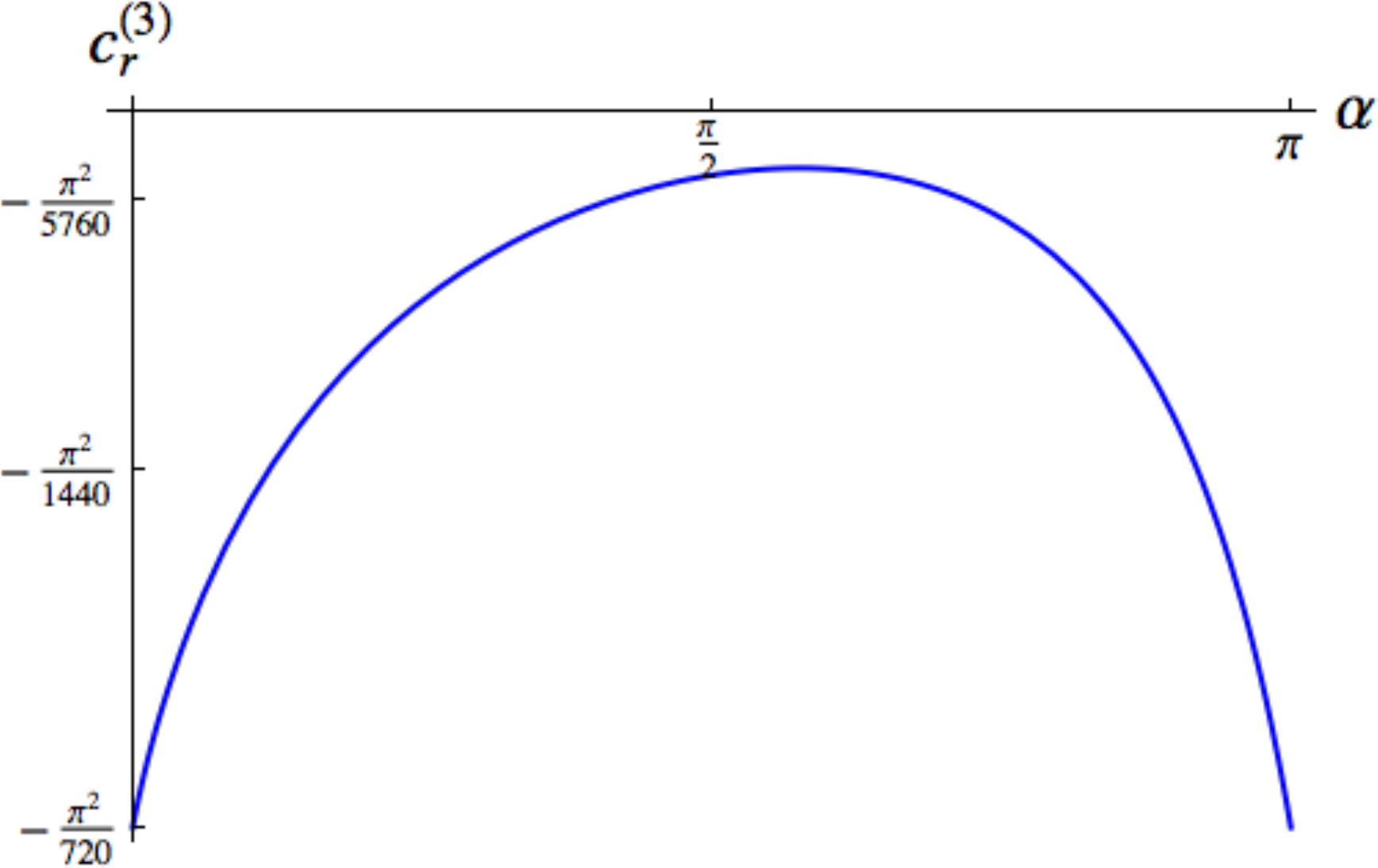}}
  \caption{\footnotesize{$\alpha$-dependence  of the ${c}^{(3)}_{r}$ coefficient of Casimir energy for Robin boundary conditions for $\alpha\in[0,\,\pi]$.}}\label{robin3d}
\end{figure}
The Casimir energy for Robin boundary conditions is displayed in Figure \ref{robin3d} which is in agreement with previous analyses \cite{romeo, farina}. Of note is that 
${c}^{(3)}_r$ is negative for all values of $\alpha\in[0,\,\pi/2]$. In other words,  the Casimir force between plates in this case is always attractive, which is in agreement with the Kenneth-Klich theorem (see below).
\begin{figure}[h]
 \centerline{\includegraphics[width=9.5cm]{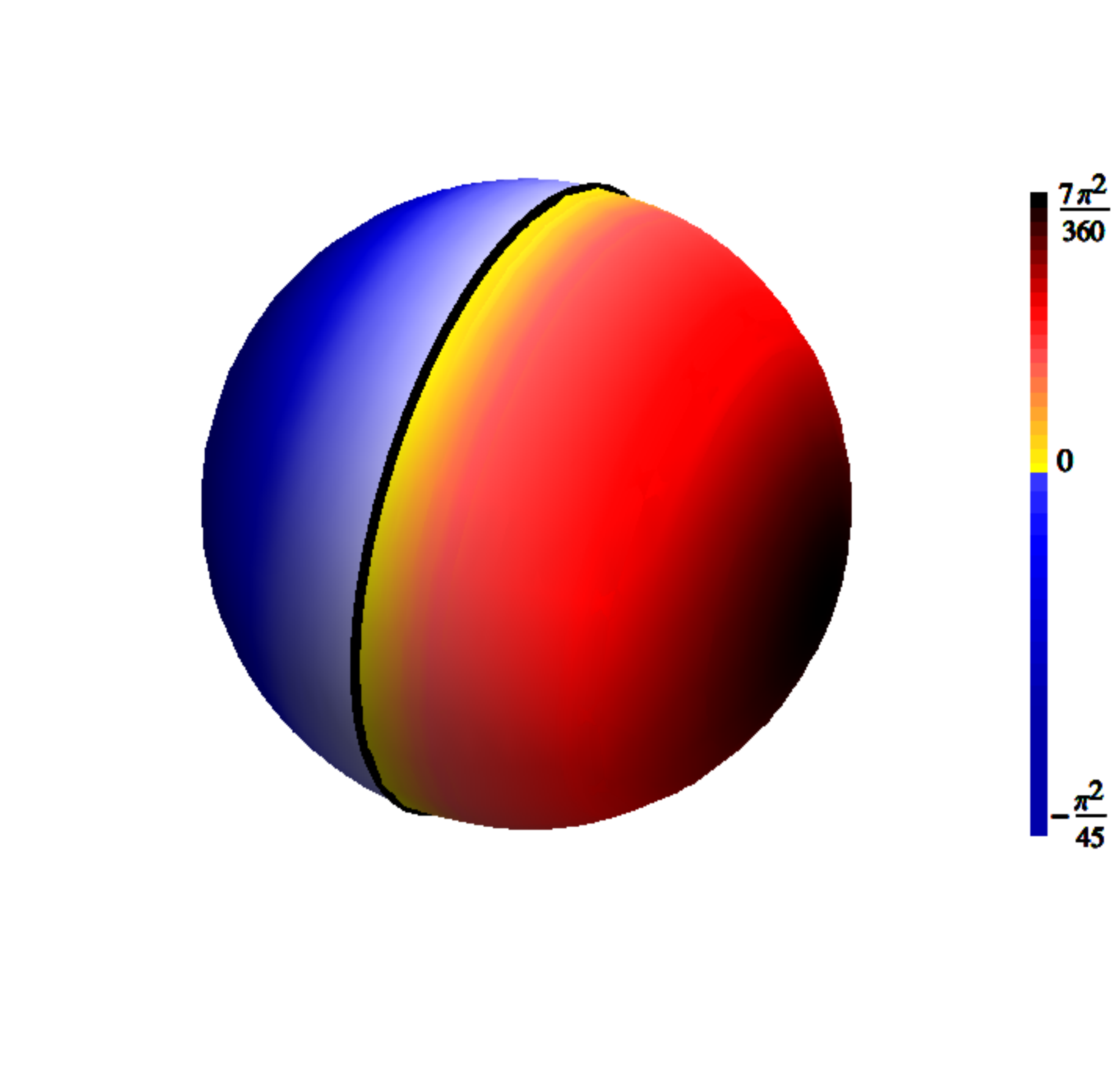}}
 \vspace{-1.5cm}
  \caption{\footnotesize{[Color inline] Variation of the  $c^{(3)}_U$ coefficient  of the Casimir energy in the consistency region $\beta=\frac{\pi}{2},\,\alpha=\frac{-\pi}{2}$ for any value of the normal vector ${\bf n}$. The black curve correspond to boundary conditions with vanishing Casimir energy and blue (red) regions  to boundary conditions with attractive (repulsive)  Casimir forces.}} \label{3dsphere}
\end{figure}

\par
\paragraph{{\rm ix)} Pauli matrices boundary conditions}  Another  case of special interest are the boundary conditions that are located at
 left and right corners of the rhombus of Figure \ref{rombo}, i.e.  boundary conditions corresponding to  points on the unit sphere $S^2$ for values $\alpha=\pm\beta=\frac{\pi}{2}$, i.e. $U_{\bf n}=\bf n\cdot \bm{\sigma}$. These two-parameter family of boundary conditions includes periodic, anti-periodic, quasi-periodic and pseudo-periodic boundary conditions. The Casimir energy  given by
 \begin{equation}
 \displaystyle{ \frac{{ E}_{{\mathrm{}}}(n_1)}{S}=\frac1{L^3}\left(-\frac{\pi ^2}{45}+\frac{(\arccos n_1)^2}{6}-\frac{(\arccos n_1)^3}{6 \pi }+\frac{(\arccos n_1)^4}{24 \pi ^2}\right)}, \label{fixed}
\end{equation}
with $ \arccos\, n_1\!\!\in\!\![0,\,2\pi]$, has 
two regimes,  attractive and repulsive, separated by a one dimensional circle of Casimirless boundary conditions (see Figure \ref{3dsphere}) given
by $\textstyle {\alpha= \beta=\textstyle\frac{\pi}{2}}$ and  
\begin{equation}
n_1=\cos\pi[{1\pm({\textstyle 1-2\sqrt{{2/15}})^\half}]}.\label{doce}
\end{equation}

\par
It is remarkable that all analytical results obtained by the spectral function method 
agree with those obtained by other methods like zeta function regularisation method.
However the  expression for the Casimir energy in terms of a contour integral of  the spectral function provides a very efficient method for   numerical  calculations of the Casimir energy in the cases where it cannot be achieved by analytic methods. 
\par
In this way we can calculate the Casimir energy for a wider class of boundary conditions using the spectral function method (\ref{cas3}). In many cases these results were previously known, and the  results obtained by the spectral function method are in perfect agreement with those found in the literature. Apart from the well-known analytic results  described above there is also agreement with the numerical simulations of cases like Robin boundary conditions where there are not analytic expressions for the Casimir energy \cite{romeo, farina}.

\begin{figure}%[htbp]
 \centerline{\includegraphics[height=8cm]{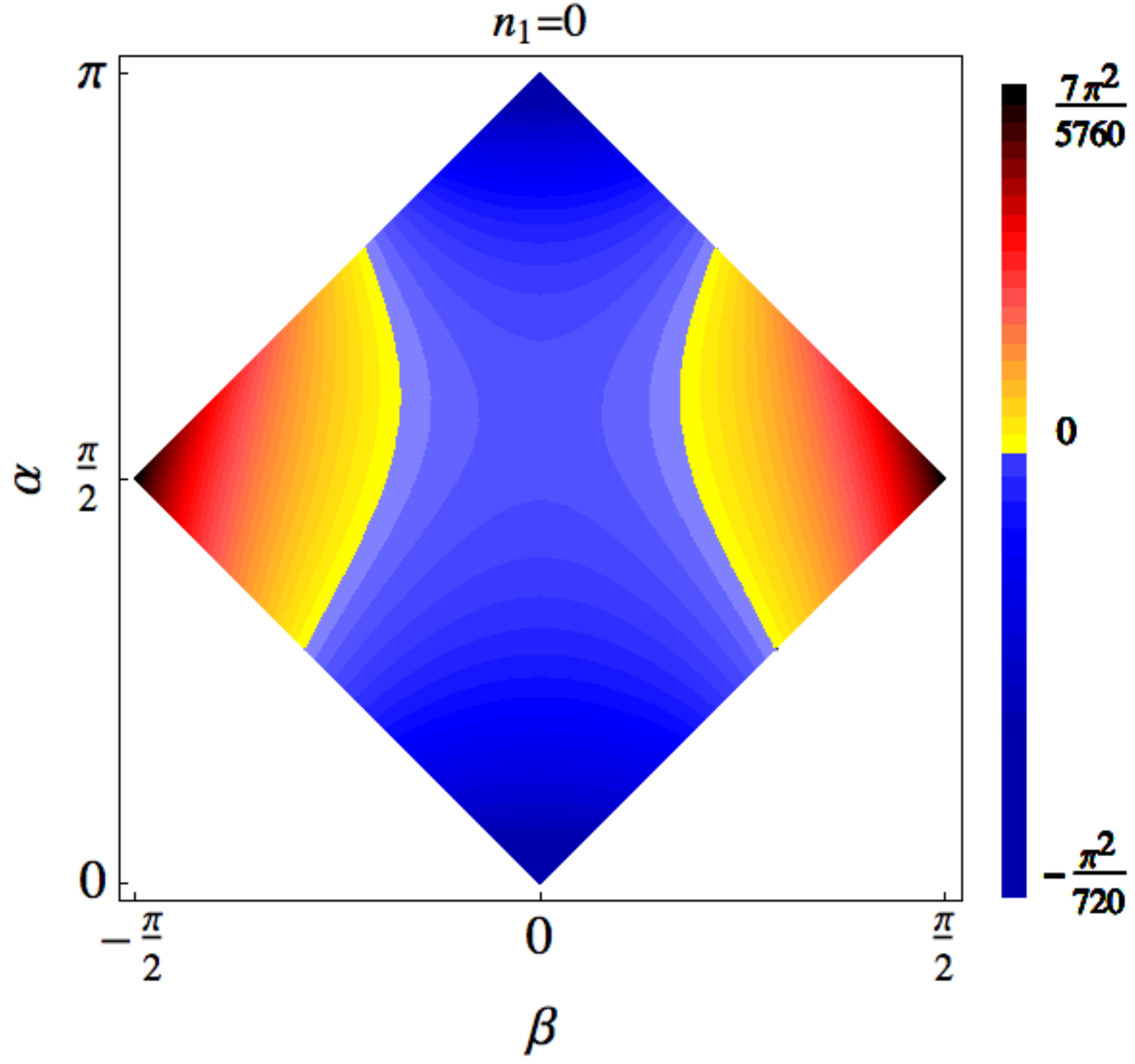}\qquad \includegraphics[height=8cm]{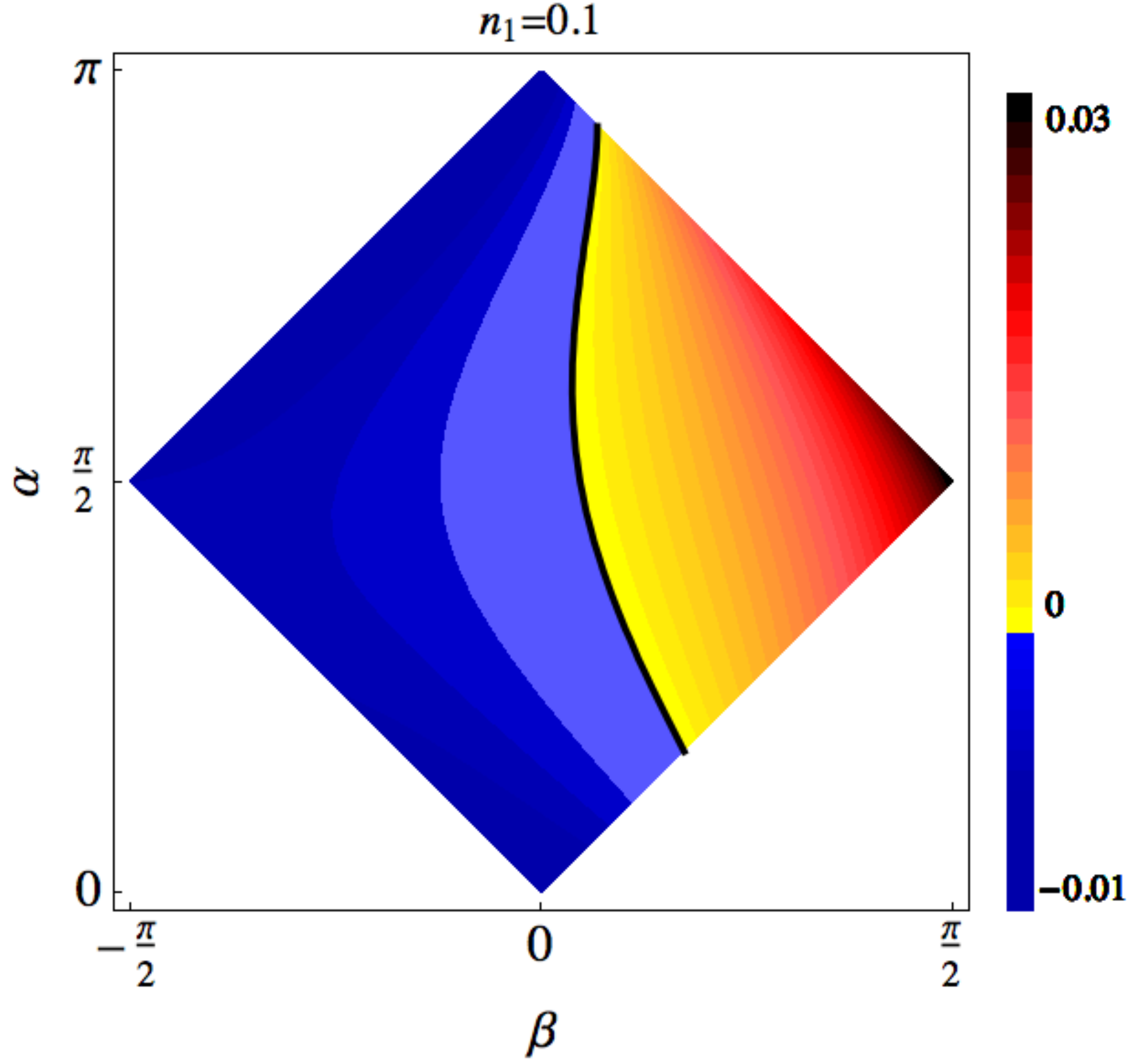}}
  \caption{\footnotesize{[color online] Variation  of the  $c^{(3)}$ coefficient of the Casimir energy in the consistency domain of boundary conditions $|\beta|<\alpha<\pi-|\beta|$, for $n_1=0$ and $n_1=0.1$. Black curves correspond to boundary conditions with vanishing Casimir energy and blue (red) regions correspond to boundary conditions with negative (positive) values of  Casimir energy.}} \label{3dcas0}
\end{figure}
\begin{figure}
  \centerline{\includegraphics[height=8cm]{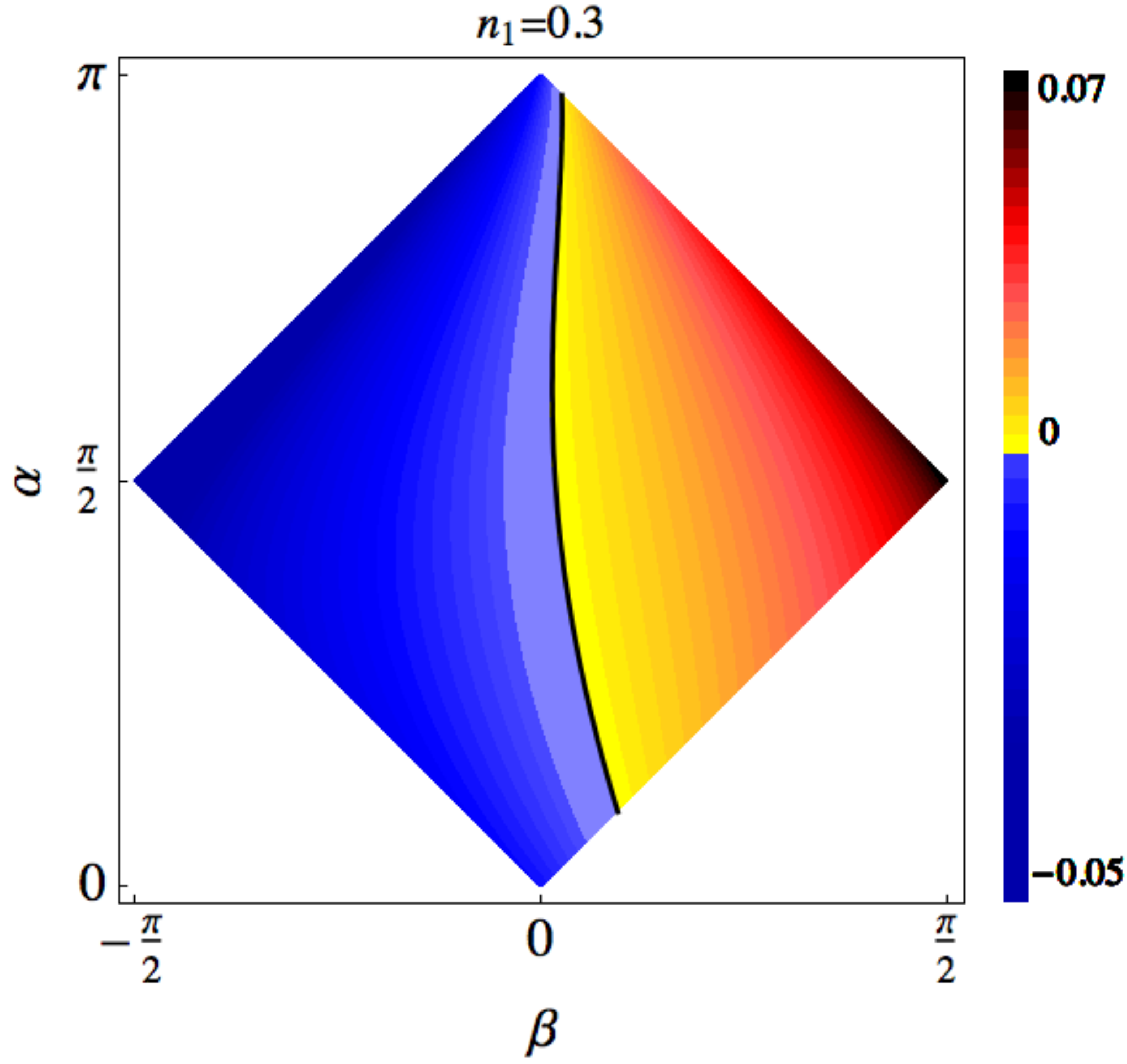} \qquad \includegraphics[height=8cm]{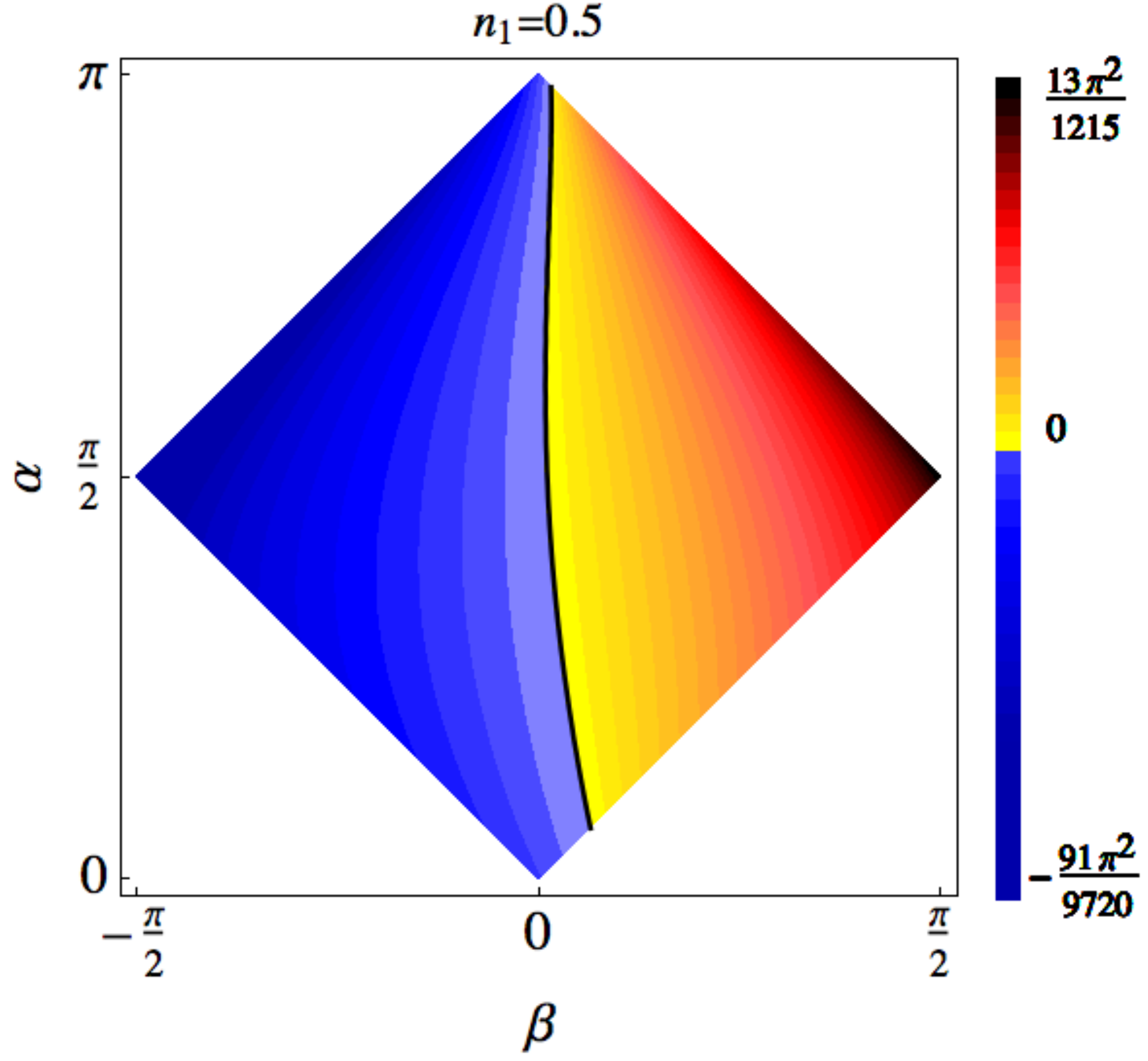}}
    \caption{\footnotesize{[color online] Variation  of the  $c^{(3)}$ coefficient of the Casimir energy in the consistency 
domain of boundary conditions $|\beta|<\alpha<\pi-|\beta|$, for $n_1=0.3$ and $n_1=0.5$. Black curves correspond to boundary conditions with vanishing Casimir energy and blue (red) regions correspond to boundary conditions with negative (positive) values of  Casimir energy.}} \label{cas3d1comp2}
\end{figure}
\begin{figure}%[htbp]
 \centerline{\includegraphics[height=8cm]{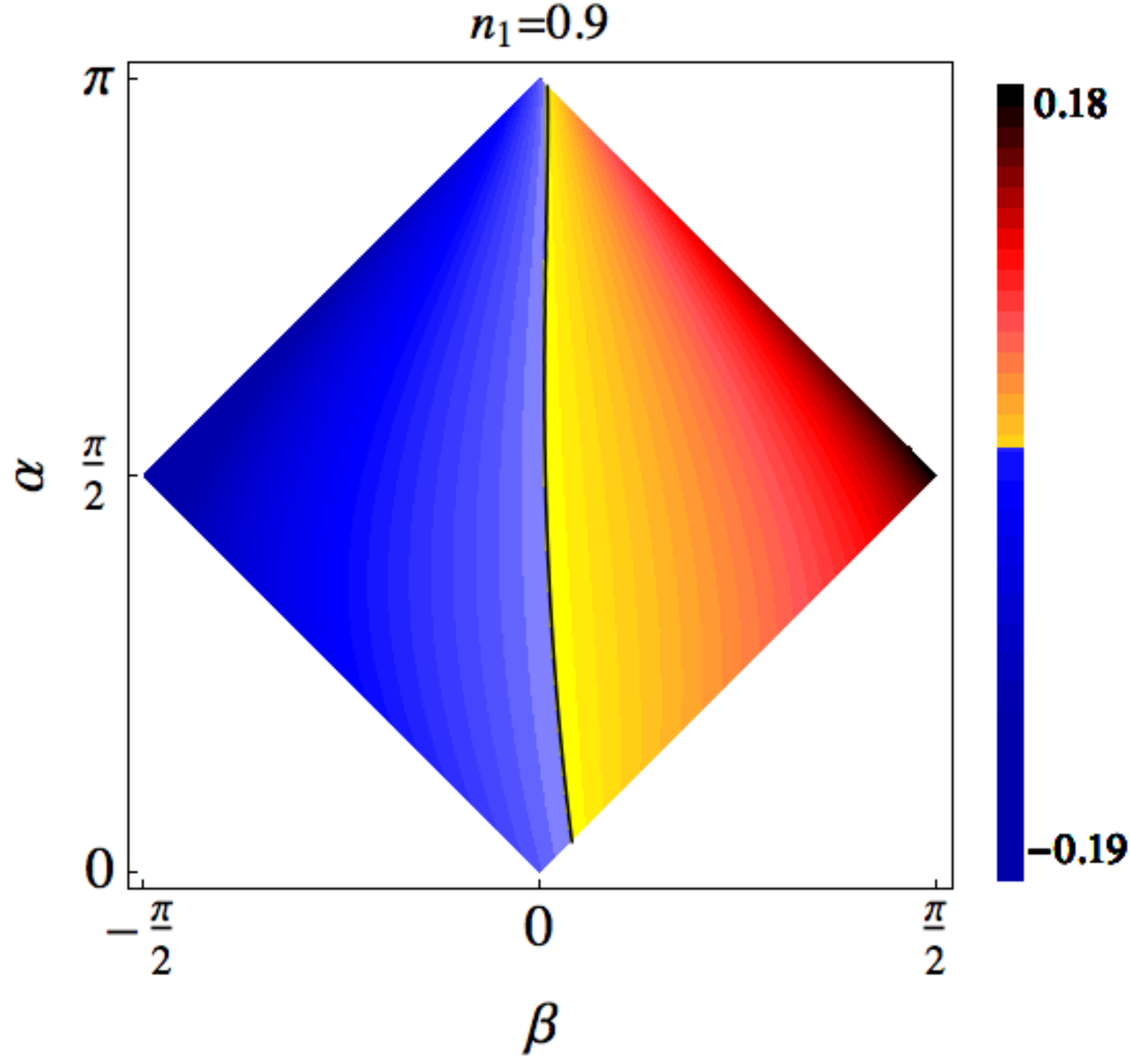}\qquad \includegraphics[height=8cm]{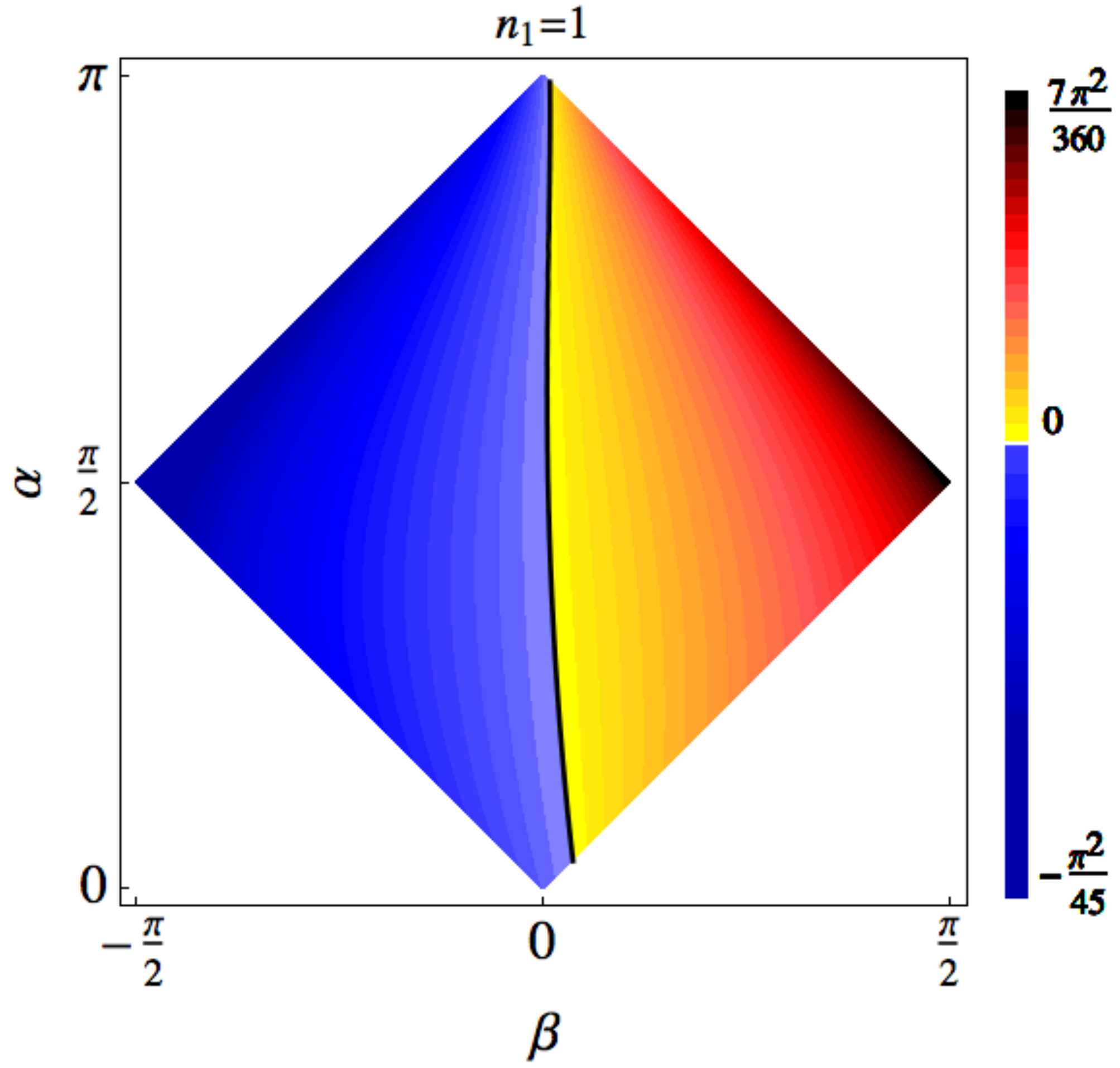}} 
  \caption{\footnotesize{[color online] Variation  of the  $c^{(3)}$ coefficient of the Casimir energy in the consistency domain of boundary conditions $|\beta|<\alpha<\pi-|\beta|$, for $n_1=0.9$ and $n_1=1$. Black curves correspond to boundary conditions with vanishing Casimir energy and blue (red) regions correspond to boundary conditions with negative (positive) values of  Casimir energy.}} \label{cas3d1comp1}
\end{figure}

%%%%%%%%%%%%%%%%%%%%%%%%%%%%%%%%%%%%%%%%%%%%%%%%%%%%%%%%%%%
\section{Casimirless Boundary conditions.}
\label{sec:5}
%%%%%%%%%%%%%%%%%%%%%%%%%%%%%%%%%%%%%%%%%%%%%%%%%%%%%%%%%%%

For generic boundary conditions there are not  analytic results but  the efficiency of the numerical analysis 
using  formula (\ref{cas3}) allows the calculation of the Casimir energy for any  boundary condition in one simple step.
The space of boundary conditions $\mpro_F$ is four dimensional, however, the Casimir energy only depends on three parameters $\alpha,\,\beta,\,n_1$, i.e. it is independent of the value of $n_2$, $n_3$ components of the unitary vector ${\bf n}=(n_1,n_1,n_3)$. Hence a global calculation  of Casimir energy reduces to the calculation of a family of planar functions on  slices of $\mpro_F$ parametrized  by the different  values of $n_1\in[-1,1]$. In each slice the Casimir energy can be represented by its contour lines and this makes the identification of the attractive and repulsive regimes easier by highlighting the curves where the Casimir energy vanishes. 
  
In this representation there are redundancies because some points in the slices $n_1$ and $-n_1$ correspond to the same boundary conditions. For this a reason we only consider positive values of $n_1$.  

%The efficiency of the numerical calculations using  formula (\ref{casimir}) allows the calculation of the Casimir energy for any $U\in\mpro_F$  in one step. 

The results show that  Casimir energy, as a function $c^{(3)}_U/L^3:\mpro_F\rightarrow\R$, has negative, positive and null values (see Figures  \ref{3dcas0}-\ref{cas3d1comp1}). These correspond to attractive, repulsive, and zero Casimir effect, i.e, there are operators $U\in\mpro_F$ which give rise to field theories without Casimir effect (zero Casimir energy). Such boundary conditions are characterised by the solutions of the equation 
%\begin{figure}[htbp]
%  \centerline{\includegraphics[height=8cm]{New1Df-n0.pdf}\qquad \includegraphics[height=8cm]{New1Df-n01.pdf}}
%  \caption{\footnotesize{[color inline]   $(\alpha,\beta)$--dependence of the  $c^{(1)}$ coefficient  of the Casimir energy in the consistency region $|\beta|<\alpha<\pi-|\beta|$, for $n_1=0$ and $n_1=0.1$. Black curves correspond to boundary conditions with vanishing Casimir energy and blue (red) regions correspond to boundary conditions with attractive (repulsive) Casimir forces.}} \label{casncero} 
%\end{figure}
%\begin{figure}[htbp]
%  \centerline{\includegraphics[height=8cm]{New1Df-n03.pdf} \quad\includegraphics[height=8cm]{New1Df-n05.pdf}}
%  \caption{\footnotesize{[Color inline]  $(\alpha,\beta)$--dependence of the  $c^{(1)}$ coefficient of the Casimir energy in the consistency region $|\beta|<\alpha<\pi-|\beta|$, for $n_1=0.3$ and $n_1=0.5$. Black curves correspond to boundary conditions with vanishing Casimir energy and blue (red) regions correspond to boundary conditions with attractive (repulsive) Casimir forces.}} \label{casextr}
%\end{figure}

\begin{equation}
  \int_0^\infty dk k^3\left[L-L_0- {}{}\frac{d}{dk}\log\left(\frac{h_U^{(L)}(ik)} {h_U^{(L_0)}(ik)}\right)\right]=0.\label{cless}
\end{equation}
in  $\mpro_F$.
 Equation (\ref{cless}) has infinite solutions in $\mpro_F$ as can be seen from Figures \ref{3dcas0}-\ref{cas3d1comp1}. Null Casimir energy subspaces are 3-dimensional and their co-dimension is 1 because they only satisfy one equation constraint (\ref{cless}).

%\begin{figure}[htbp]
%  \centerline{\includegraphics[height=8cm]{New1Df-n09.pdf} \qquad\includegraphics[height=8cm]{New1Df-n1.pdf}}
%  \caption{\footnotesize{[color inline] $(\alpha,\beta)$-dependence of the  $c^{(1)}$ coefficient of the Casimir energy in the consistency region $|\beta|<\alpha<\pi-|\beta|$, for $n_1=0.9$ and $n_1=1$.  Black curves correspond boundary conditions with vanishing Casimir energy and blue (red) regions correspond to boundary conditions with attractive (repulsive) Casimir forces.}} \label{casnmas}
%\end{figure}

\par
The most relevant property of the 3-dimensional subspace of Casimirless boundary conditions is that it  splits { the space of physical} boundary conditions $\mpro_F$ into two disjoint subsets: one  containing all boundary conditions that generate  repulsive  Casimir force and the other  containing those with attractive Casimir force. In other words the existence of Casimirless boundary conditions highlights the transition between both types of regimes. Although the  subspace of  boundary conditions with vanishing Casimir energy  is  connected, in some slices  the intersection of the subspace defined by (\ref{cless}) has two connected components curves while for others it has only one connected component.

The numerical results show that the minimum of  Casimir energy is obtained with  periodic boundary conditions, which means that these boundary conditions generate the strongest attractive Casimir force between the plates. In the same way we find that the maximum value of  Casimir energy is obtained with anti-periodic boundary conditions, which means that these boundary conditions 
generate the strongest repulsive Casimir force.

Using the numeric calculations above we can verify that the Kenneth-Klich theorem \cite{KK} also holds in 3+1 dimensions. The theorem states that the Casimir force between two identical bodies is always attractive. It is implicit in the assumptions of the theorem that the boundary conditions introduced by  the two bodies are each other independent. The only conditions that satisfy this property are those with $n_1=n_2=0, n_3=1$, i.e.
\begin{equation}
  U(\alpha,\beta,(0,0,1))=\e^{i\alpha}\left(\cos\beta\I +i\sigma_3\sin\beta\,{}\right)=\left(\begin{array}{cc}
\e^{i(\alpha+\beta)}&  0 \\ 0  &\e^{i(\alpha-\beta)} \\ 
\end{array}\right).\label{uindep-bodies}
\end{equation}
Boundary conditions which modelling identical bodies require $\beta=0$ and in this case we have identical Robin boundary conditions
with $n_1=0$ which corresponds to boundary conditions sitting on the vertical line connecting the Dirichlet and Neumann corners of the rhombus in Figure \ref{rombo}. In fact the above results show that the same behaviour hold for bodies with slightly different boundary conditions. This follows from the continuity of the Casimir energy in the space of boundary conditions $\mpro_F$. The repulsive behaviour requires in general a rather different boundary conditions for the two plates, e.g. Zaremba  boundary conditions ($\alpha=\pm \beta=\frac{\pi}2$, i.e. $U=\pm \sigma_3$) located on the left (right) corners of the rhombus which correspond to two plates, one with Neumann boundary conditions whereas the other has Dirichlet boundary conditions.

%The calculations are in perfect agreement with the expected  behaviour of Casimir energy implied by the 
%Kenneth-Klich theorem  in 3+1 dimensions, i.e.  the  Casimir force between two identical bodies with identical 
%boundary conditions is always attractive.
%If the two  plates are identical and have the same boundary conditions  $\beta=n_1=0$ and in that case 
%
% In general we also expect that, due the to
%continuity of  Casimir energy in the space of boundary conditions $\mpro_F$, the attractive regime extends
%for bodies with slightly different Robin boundary conditions as shown by the numerical results (see Fig.  (\ref{3dcas0})).
%As in lower dimensional field theories, the repulsive behaviour of Casimir force 
% appears for bodies with very different boundary conditions, like Zaremba
% boundary conditions.
 
 The  same behaviour appears for higher dimensional D+1 field theories. In these cases the Casimir energy for identical plates is 
 given by
\begin{equation}
  \frac{c^{(2n+1)}_r}{L^{2n+1}}=\nonumber \displaystyle{
  \frac{(-1)^{n}(4\pi)^{-\frac{2n+1}{2}}\Gamma\left(-\frac{2n+1}{2}\right)\, L_0^{2n+1}}{\pi   (L^{2n+1}-L_0^{2n+1})} \int_0^\infty dk\ k^{2n+1}}
 \left[L^{}-L_0^{}-\frac{d}{dk}\log\left(\frac{h_r^{(L)}(ik)} {h_r^{(L_0)}(ik)}\right)\right]\label{casimir3r}
\end{equation}
for D=2n+1 odd dimensions, 
or 
\begin{equation}
  \frac{c^{(2n)}_r}{L^{2n}}={\displaystyle - \frac{{ }(4{\pi})^{-n}\, L_0^{2n} }{  \Gamma(n+1) (L^{2n}\!-\!L_0^{2n})}}
  \int_0^\infty dk\, k^{2n} 
  \left[L-L_0-\frac{d}{dk}\log\left(\frac{h_r^{(L)}(ik)} {h_r^{(L_0)}(ik)}\right)\right]\label{casimir2nr}
\end{equation}
for  D=2n even dimensions.
Since $ (-1)^{n}\Gamma\left(-\frac{2n+1}{2}\right)$ is always negative it is sufficient to prove that the integrand is
in both cases is a positive function of $k$.
Boundary conditions introduced by two independent identical bodies are given by (\ref{uindep-bodies}) with $\beta=0$. The associated spectral function is
\begin{equation}{
h_r^{(L)}(ik)=2{\rm e}^{i \alpha}\left({\left( k \cos \frac{\alpha}{2} + \sin \frac{\alpha}{2}\right)^2{\rm e}^{kL}- \left(k \cos \frac{\alpha}{2} - \sin \frac{\alpha}{2}\right)^2{\rm e}^{-kL}}\right),}
\end{equation}
and 
\begin{equation}
\frac{h_r^{(L)\prime}(ik)}{h_r^{(L)}(ik)}= \frac{ 4 k \cos^2 \frac{\alpha}{2}\sinh{kL} + 2\sin \alpha\cosh{kL}+ L(k \cos \frac{\alpha}{2} + \sin \frac{\alpha}{2})^2{\rm e}^{kL}+ L (k \cos \frac{\alpha}{2} - \sin \frac{\alpha}{2})^2{\rm e}^{-kL}}
{   (k \cos \frac{\alpha}{2} + \sin \frac{\alpha}{2})^2{\rm e}^{kL}- (k \cos \frac{\alpha}{2} - \sin \frac{\alpha}{2})^2{\rm e}^{-kL}}.
\end{equation}
When $L>L_0$  the inequality
\begin{equation}
{h_r^{(L)\prime}(ik)/{h_r^{(L)}(ik)}-h_r^{(L_0)\prime}(ik)/{h_r^{(L_0)}(ik)}< L-L_0 }\label{ineq}
\end{equation}
provides the necessary bound which ensures that the integral  is always positive. Thus in any dimension  the Casimir energy is always negative 
$$
c_r^D <0
$$
as required by Kenneth-Klich theorem.

%%%%%%%%%%%%%%%%%%%%%%%%%%%%%%%%%%%%%%%%%%%%%%%%%%%%%%%%%%%
\section{Casimir energy  in 2-dimensions.}
\label{sec:7}
%%%%%%%%%%%%%%%%%%%%%%%%%%%%%%%%%%%%%%%%%%%%%%%%%%%%%%%%%%%

Two dimensional systems have acquired recent interest since the appearance of new  materials like graphene and  new physical effects which are specific of two dimensional systems like the quantum Hall effects. On the other hand as we have shown the calculation of
Casimir effect presents some subtleties in even dimensional spaces.
In the $D=2$ case  the integral for transverse modes apparently presents  logarithmic divergences for all the orders in the power expansion of parameter $\epsilon$. Some of them have been analyzed in the literature, see references \cite{bordag01,elizalde89,elizb,ambjorn83,milton}. However, it is remarkable that 
these divergences disappear, as we have shown in the preceding section, giving rise to a finite univocally defined Casimir energy  and a finite  Casimir pressure on the plates proportional to the  cubic power of the inverse  distance between plates. The  Casimir energy  in this case is given by
\begin{equation}
  \frac{c^{(2)}_U}{L^2} =-{1\over 4\pi}\frac{L_0^2}{L^2-L_0^2} \int_0^\infty dk\ k^2
  \left[L-L_0-\frac{d}{dk} \log\left(\frac{h_U^{(L)}(ik)}{h_U^{(L_0)}(ik)} \right)\right].\label{delta2}
\end{equation}
In some cases the integral (\ref{delta2}) can be analytically evaluated, e.g.  for Dirichlet/Neumann boundary conditions we get
 $$c^{(2)}_d=c^{(2)}_n=-\frac{\zeta(3)}{8 \pi},$$
%whereas for periodic boundary conditions we have
%$$c^{(2)}_p=-\frac{\zeta(3)}{ \pi},$$
%and
%$$c^{(2)}_{ap}=\frac{3\zeta(3)}{4 \pi}$$
%for anti-periodic, and
%$$c^{(2)}_z=\frac{3\zeta(3)}{32 \pi}$$for
% Zaremba boundary conditions. 
whereas we have
$$c^{(2)}_p=-\frac{\zeta(3)}{ \pi},\  c^{(2)}_{ap}=\frac{3\zeta(3)}{4 \pi}, \  c^{(2)}_z=\frac{3\zeta(3)}{32 \pi},$$
for periodic, anti-periodic and Zaremba boundary conditions. The results 
are finite and in agreement with those obtained using other methods such as zeta function regularisation method \cite{bordag01,elizalde89,elizb,ambjorn83,milton}. 

We also obtain analytic expressions for the Casimir energy for quasi-periodic (\ref{cuasiu}) and pseudo-periodic (\ref{pseudou}) 
boundary conditions
\begin{eqnarray}
  c^{(2)}_{qp}&=&-{1\over 2\pi}\left(\li_3\left(-i\e^{i\alpha}\right)+\li_3\left(i\e^{-i\alpha}\right)\right); \quad\alpha\in\left[-\pi/2,\pi/2\right]
  \label{qpqp} \\
  c^{(2)}_{pp}&=&-{1\over 2\pi}\left(\li_3\left(\e^{i\alpha}\right)+\li_3\left(\e^{-i\alpha}\right)\right); \quad\alpha\in\left[-\pi,\pi\right],
  \label{pppp}
\end{eqnarray}
although in these cases the $\alpha$-dependence of the Casimir energy is not polynomial unlike in 3+1 dimensions (see section \ref{sec:8}).

The results  explicitly show  the universal character of the Casimir energy, even for $2+1$ dimensional space-times where some authors suggested the presence of logarithmic divergencies which would make the Casimir phenomenon dependent on the regularisation method and the renormalisation scheme. We have demonstrated by using a heat kernel regularisation  the absence of these divergencies and proved the universal character of the Casimir energy between plates for any  even dimensional space.

 In 2+1 dimensions there are also boundary conditions that generate  attractive and repulsive Casimir effects
 as in  $3+1$ dimensions. In the interface between both regimes there are boundary conditions that do not generate any Casimir force. Those boundary conditions of $\mpro_F$ with vanishing Casimir energy  are characterised as the solutions of equation 
\begin{equation}
\int_0^\infty k^2\left[L-L_0-\frac{d}{dk} \log\left(\frac{h_U^{(L)}(ik)}{h_U^{(L_0)}(ik)} \right)\right]\,dk=0,
\end{equation}
in $\mpro_F$.

%\begin{figure}[htbp]
%\centerline{\includegraphics[height=5.5cm]{New2Df-qp.pdf}}
%\caption{\footnotesize{Variation of the $c^{(2)}_U$ coefficient  of  Casimir energy as a function of $\alpha\in[-\pi/2,\,\pi/2]$ for quasi-periodic boundary conditions }} \label{delta2cp}
%\end{figure}

\begin{figure}[h]
\centerline{\includegraphics[height=5.5cm]{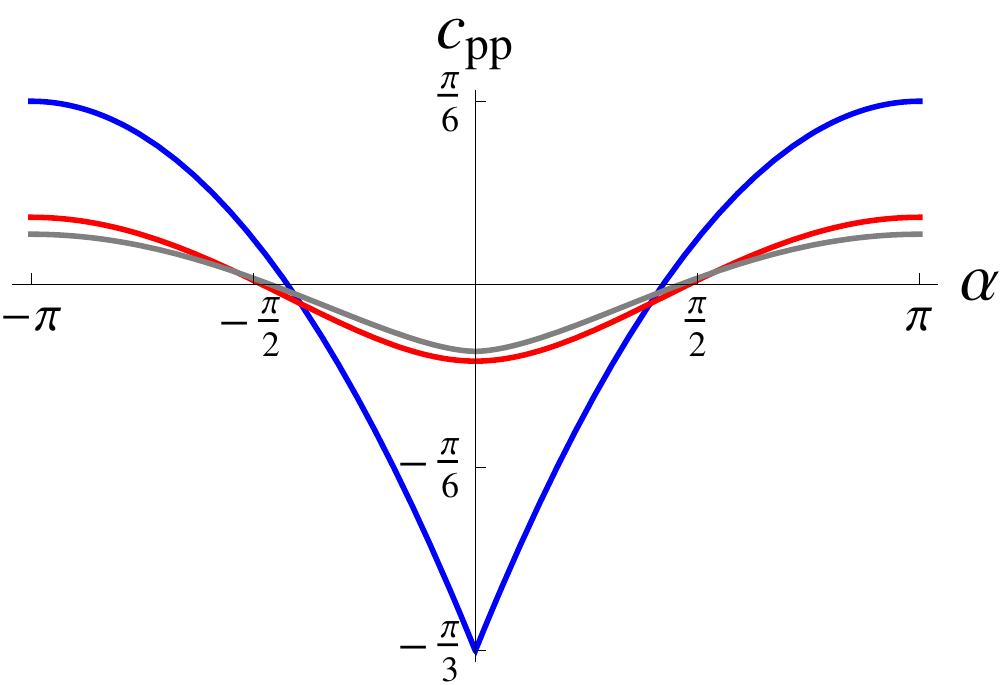}}
\caption{\footnotesize{Variation of the   $c^{(D)}_{pp}$ coefficient  of the Casimir energy  as a function of  $\alpha\in[-\pi,\,\pi]$ for psedoperiodic boundary conditions in $D=1$ (blue), $D=2$ (gray),$D=3$ (red).}  The (weak) first order transition associated to the cusp singularity only appears in  one-dimensional systems.} \label{mergedpp}
\end{figure}

In particular,  Casimirless boundary conditions arise in the one-parameter families  of quasi-periodic and pseudo-periodic boundary conditions where the Casimir energies given by  (\ref{qpqp}) and   (\ref{pppp})  point out the 
existence of three values of  the $\alpha$ parameter  for which the coefficient $c^{(2)}_U$ vanishes (see Figure \ref{mergedpp}). They correspond to Casimirless boundary conditions.
%\begin{figure}[htbp]
%\centerline{\includegraphics[height=5.5cm]{New2Df-pp.pdf}}
%\caption{\footnotesize{Variation of the   $c^{(2)}_U$ coefficient  of the Casimir energy as a function of  $\alpha\in[-\pi,\,\pi]$ for pseudo-periodic boundary conditions.}} \label{delta2pp}
%\end{figure}

%Notice that in these cases the $\alpha$-dependence of the Casimir energy is not polynomial unlike in 3+1 dimensions (see section \ref{sec:8}). 

\begin{figure}[htbp]
\centerline{\includegraphics[height=5.5cm]{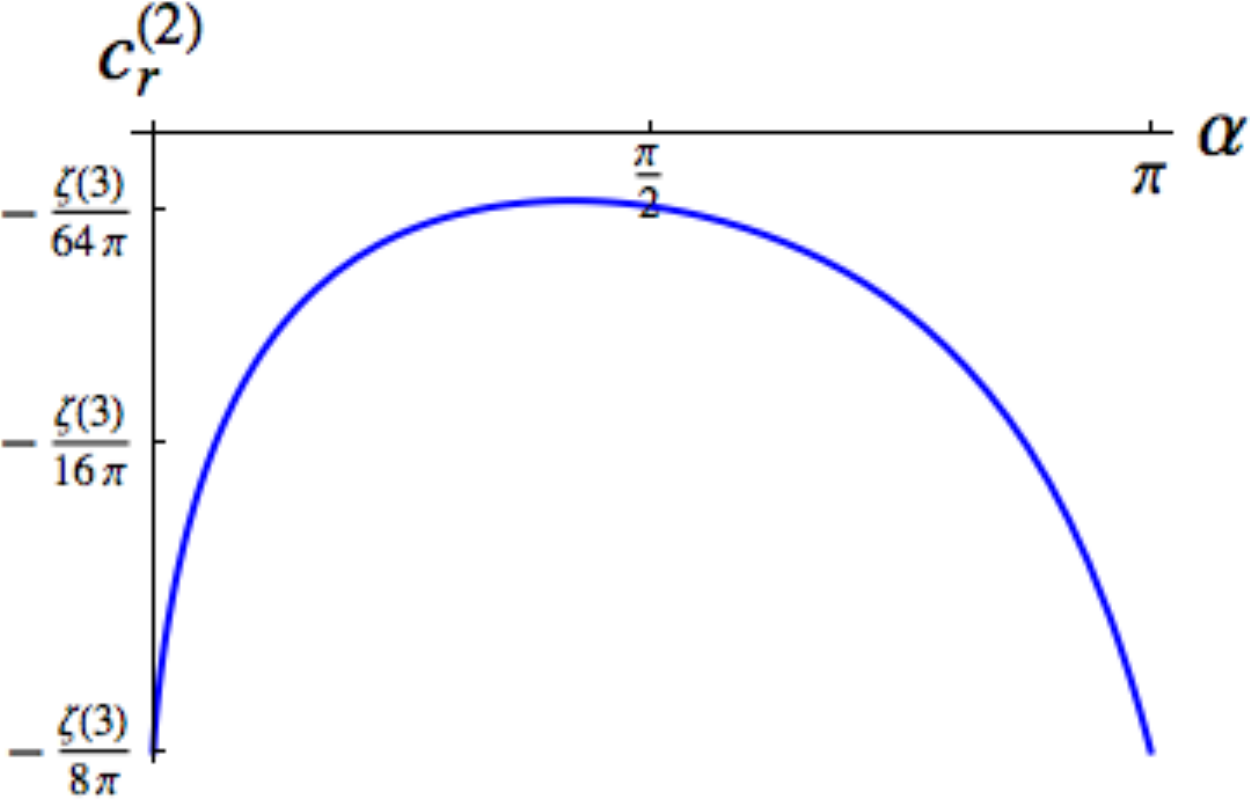}}
\caption{\footnotesize{Variation of the   $c^{(2)}_U$ coefficient  of the Casimir energy as a function of  $\alpha\in[-\pi,\,\pi]$ for Robin boundary conditions.}} \label{delta2r}
\end{figure}

In contrast, for Robin boundary conditions the numerical result is displayed in Fig. \ref{delta2r} show that the Casimir energy is 
always  negative for any value of $\alpha\in[0,\,\pi/2]$. In other words, in this case the Casimir force between plates is always attractive.% which is in agreement with the Kenneth-Klich theorem as we have anticipated at the end of previous section.

\begin{figure}[htbp]
\centerline{\includegraphics[height=8cm]{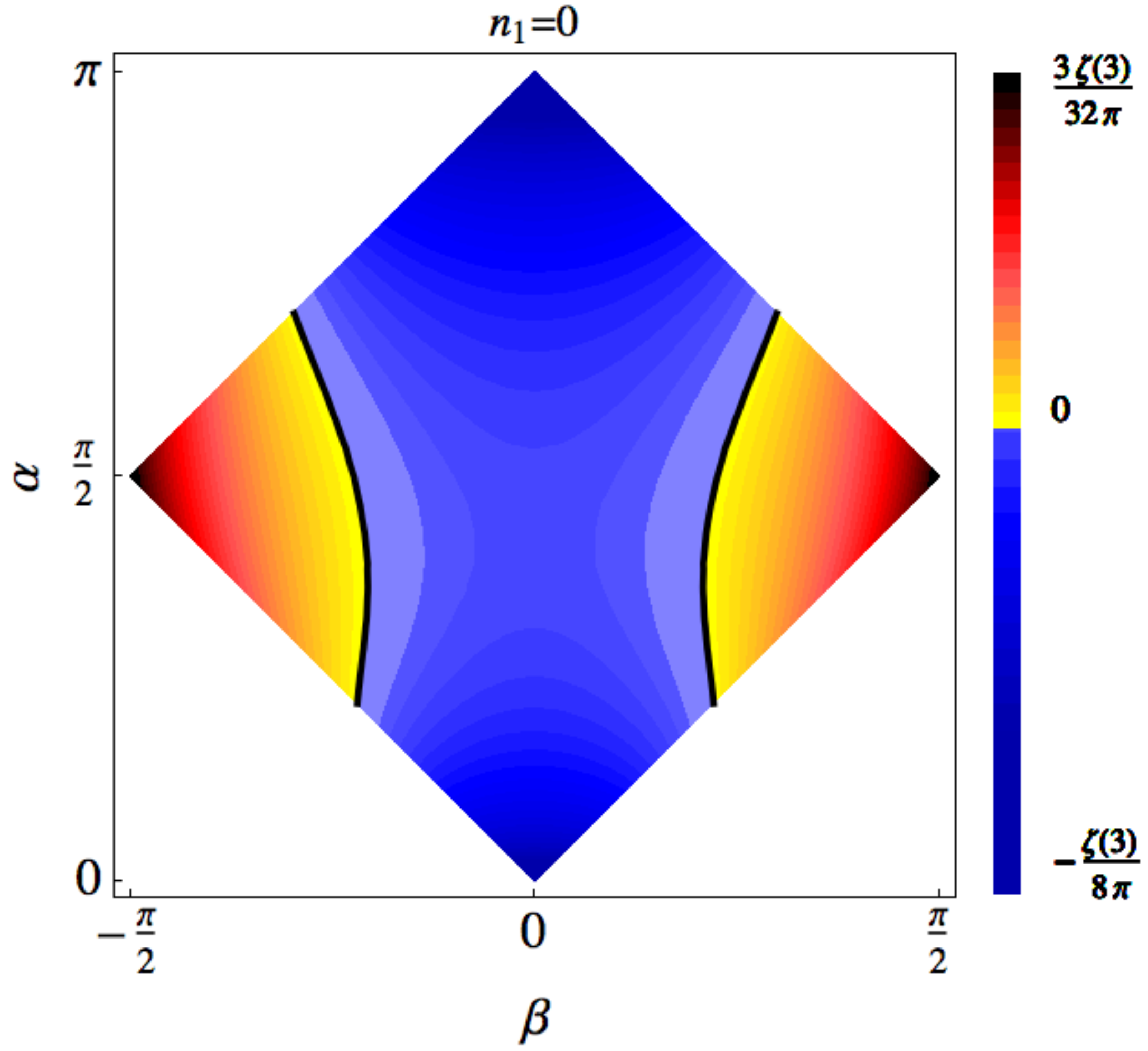}\qquad \includegraphics[height=8cm]{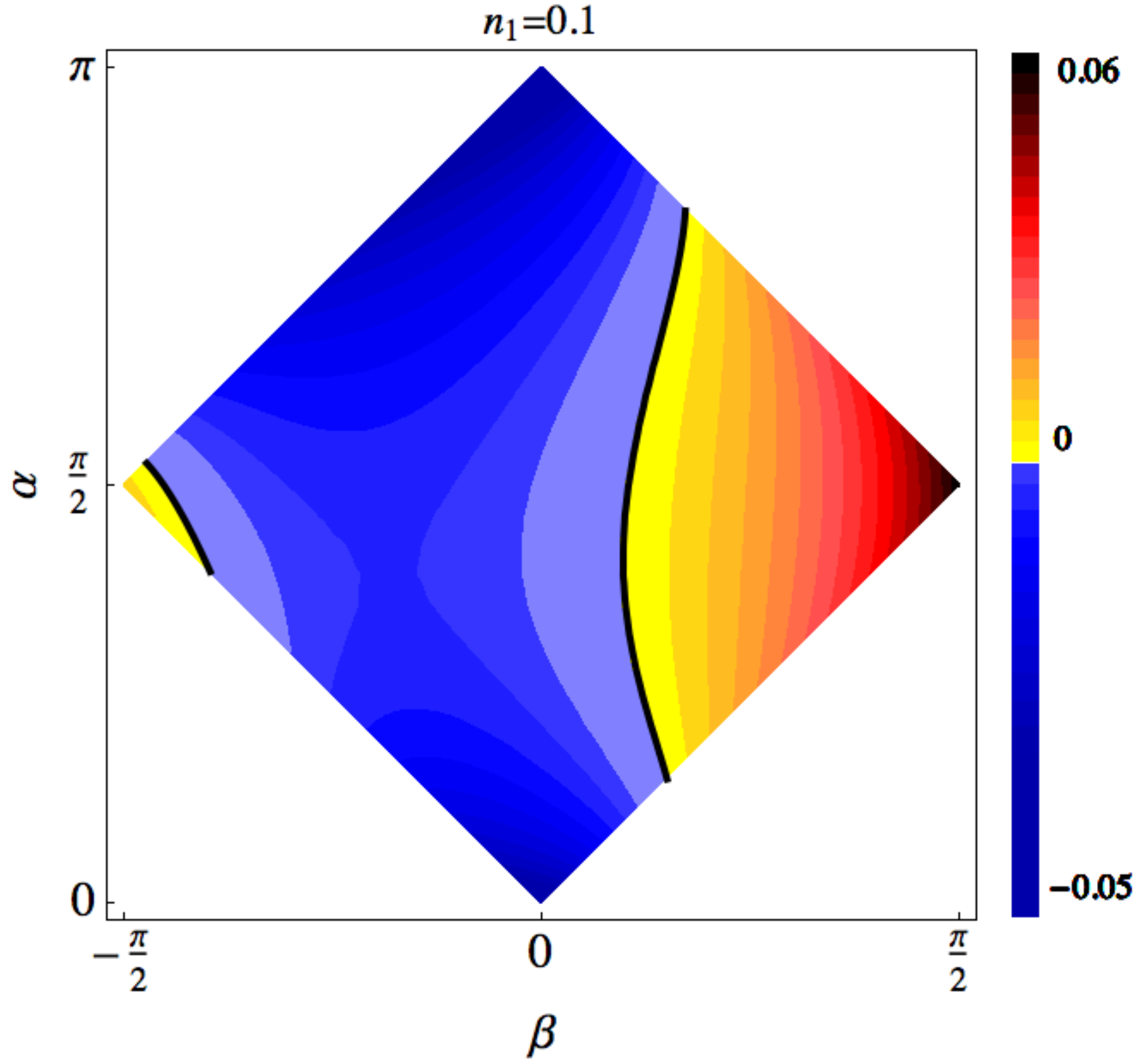}}
\caption{\footnotesize{ [color on line]  Variation  of the  $c^{(2)}$ coefficient of the Casimir energy  in the consistency domain of boundary conditions $|\beta|<\alpha<\pi-|\beta|$ for $n_1=0$ and $n_1=0.1$. Black curves correspond to boundary conditions with vanishing Casimir energy and blue (red) regions correspond to boundary conditions with attractive (repulsive)  Casimir forces.}} \label{deltados0}
\end{figure}
\begin{figure}[htbp]
\centerline{\includegraphics[height=8cm]{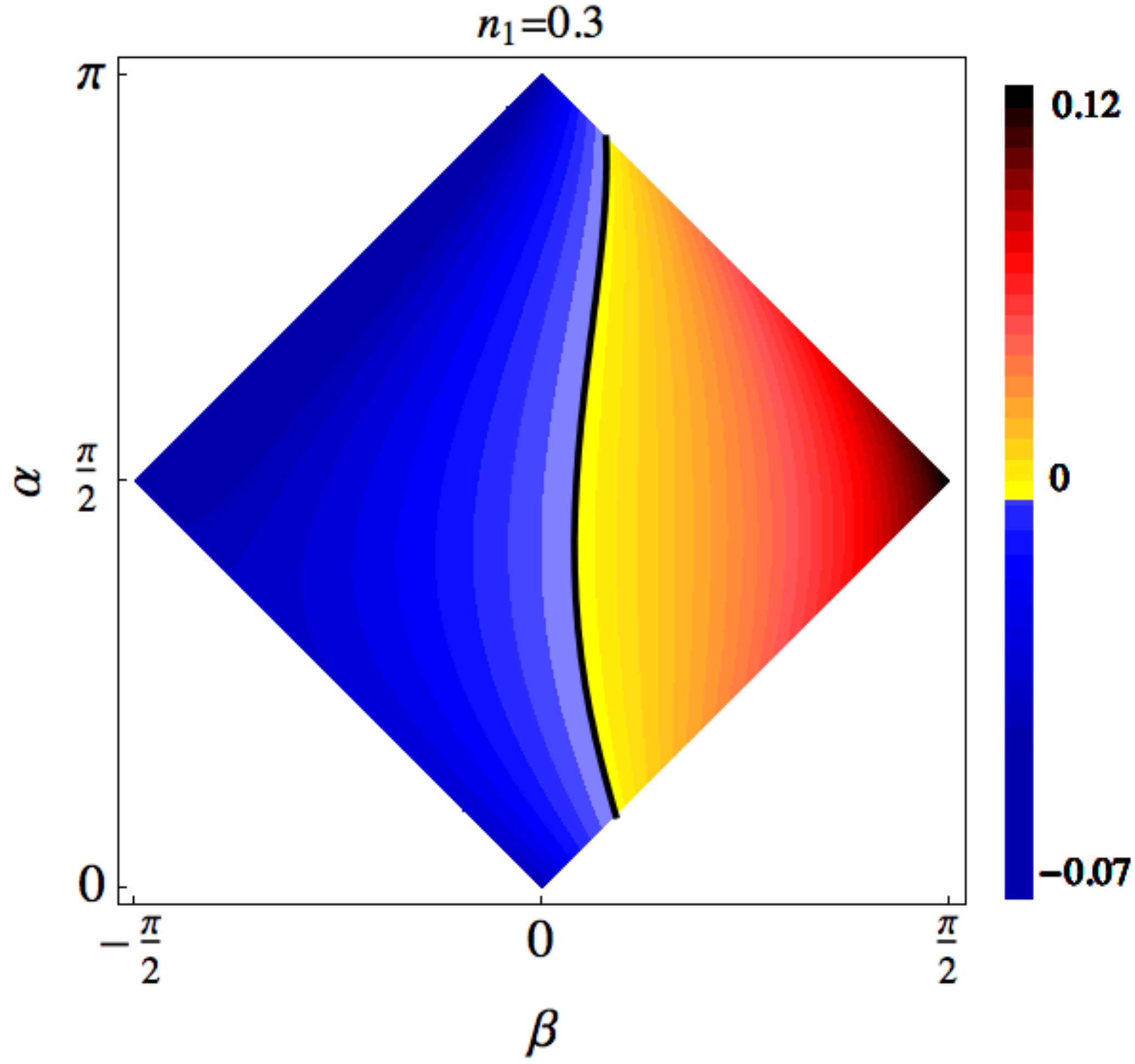} \qquad \includegraphics[height=8cm]{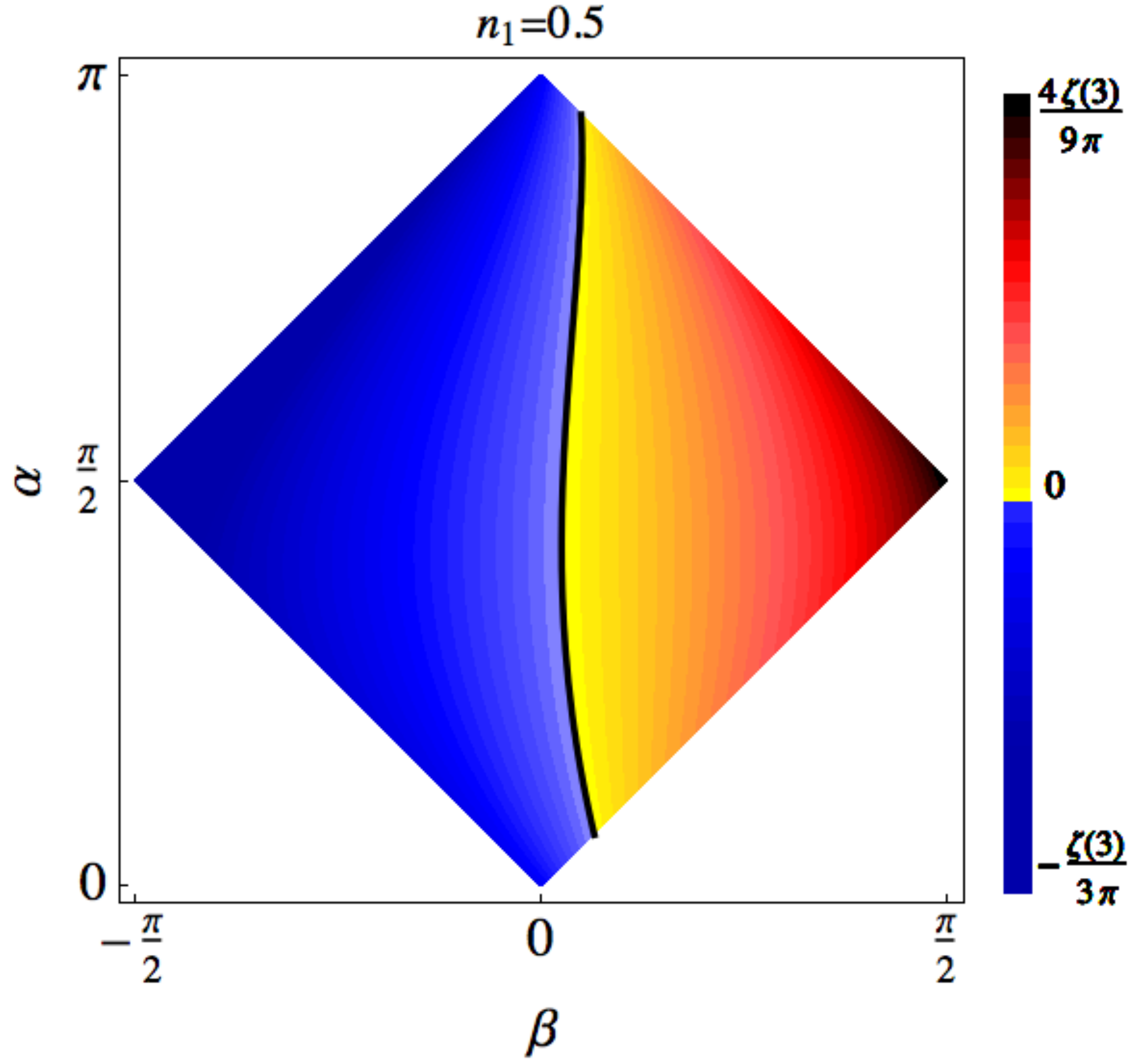}}
\caption{\footnotesize{[color on line] Variation  of the  $c^{(2)}$ coefficient of the Casimir energy in the consistency domain of boundary conditions $|\beta|<\alpha<\pi-|\beta|$ for $n_1=0.3$ and $n_1=\pm0.5$. Black curves correspond to boundary conditions with vanishing Casimir energy and blue (red) regions correspond to boundary conditions with attractive (repulsive)  Casimir forces.}} \label{deltadosp1}
\end{figure}
\begin{figure}[htbp]
\centerline{\includegraphics[height=8cm]{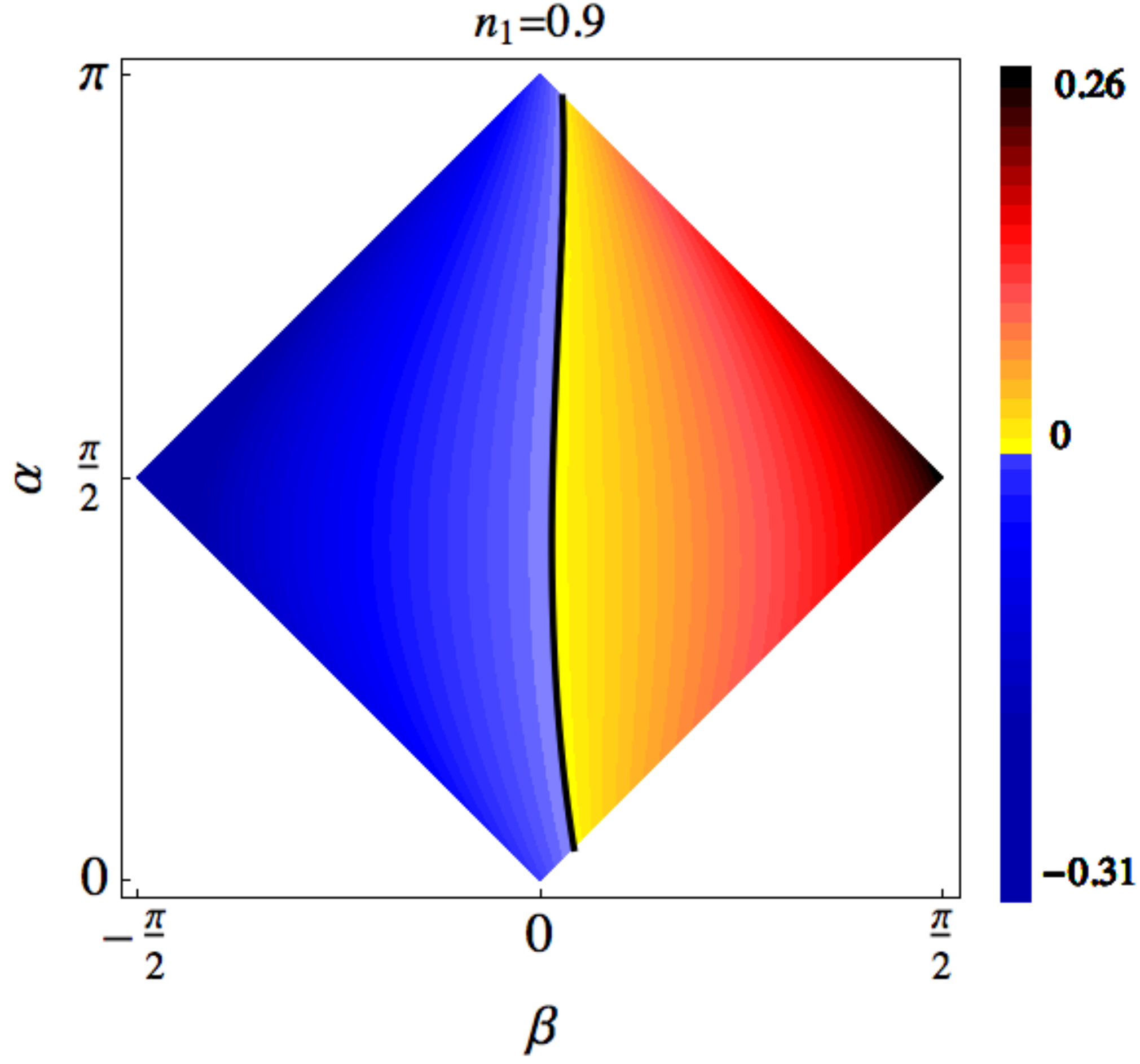} \quad\includegraphics[height=8cm]{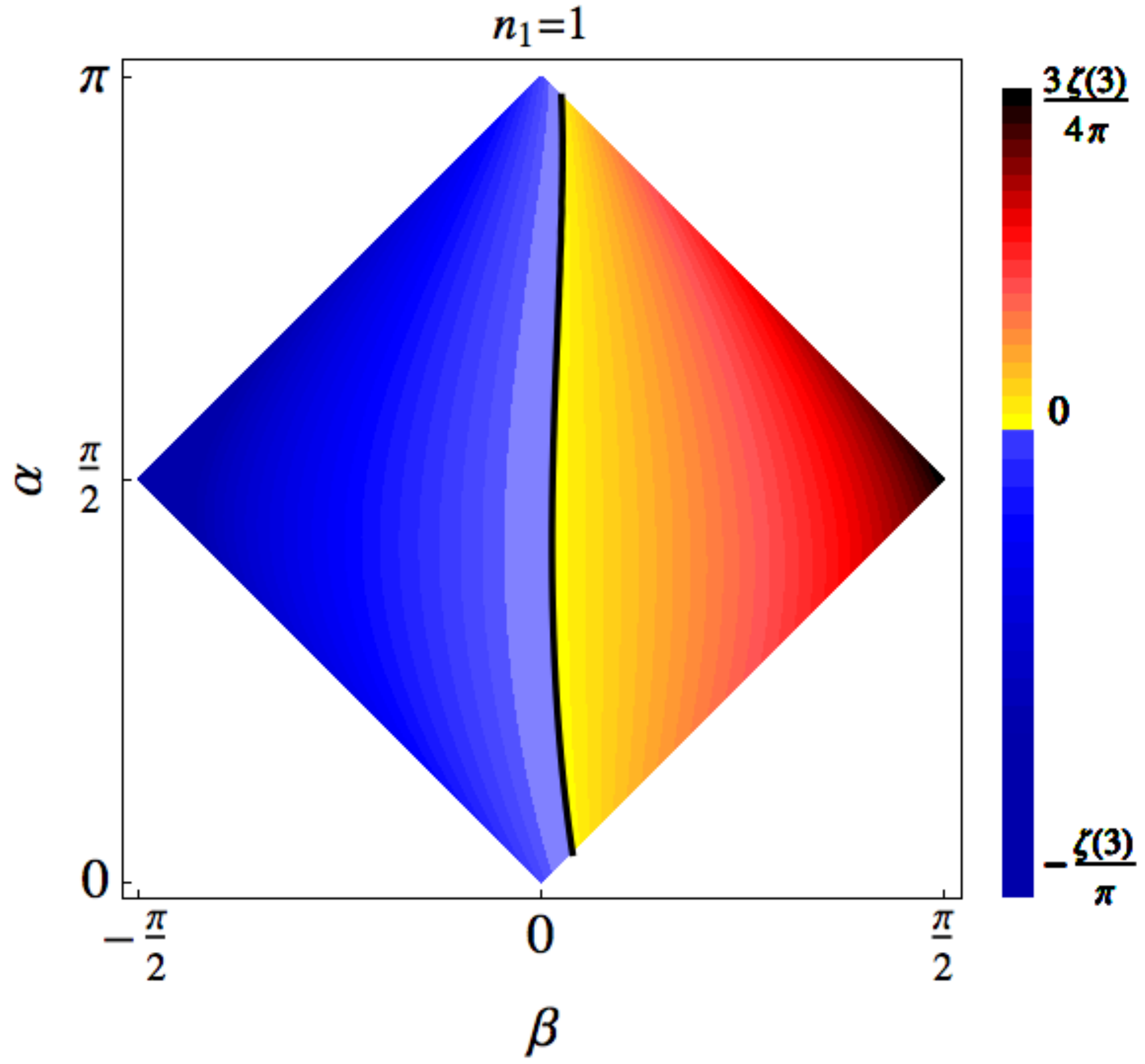}}
\caption{\footnotesize{[Color inline] Variation  of the  $c^{(2)}$ coefficient of the Casimir energy in the consistency  domain of boundary conditions $|\beta|<\alpha<\pi-|\beta|$ for $n_1=0.9$ and $n_1=1$. Black curves correspond to boundary conditions with vanishing Casimir energy and blue (red) regions  to boundary conditions with attractive (repulsive)  Casimir forces.}} \label{deltados1}
\end{figure}

Finally, using equation (\ref{delta2}) the behavior of $c^{(2)}_U$ can be numerically evaluated in  the whole domain $\mpro_F$ of consistent boundary conditions. Figures \ref{deltados1}, \ref{deltadosp1} and \ref{deltados0} show contour plots of $c^{(2)}_U$ for different values of parameter $n_1$.

Again it is explicitly shown  that for any value of $n_1$ there are curves of boundary conditions with vanishing Casimir energy  (thick lines). For the rest of boundary conditions in $\mpro_F$ the Casimir energy can take positive  and negative  values, which correspond to repulsive or attractive Casimir forces between the plates.

The  results  also confirm the behaviour inferred from the Kenneth-Klich theorem  in 2+1 dimensions. 
Indeed the Casimir force between two identical wires in 2+1 dimensions  is always attractive  as
corresponds to the case of  two identical Robin boundary conditions.
This  behaviour of the Casimir force is not exclusive of bodies with identical Robin boundary conditions, i.e. $\beta=n_1=0$
but also  for bodies with slightly different Robin boundary conditions. 
 This follows from the continuity of the Casimir energy in the space of boundary conditions $\mpro_F$ and
demonstrates that  the repulsive character of Casimir force can only 
 appear for bodies with very different boundary conditions e.g. mixed Dirichlet-Neumann conditions as in   Zaremba
 boundary conditions.

%%%%%%%%%%%%%%%%%%%%%%%%%%%%%%%%%%%%%%%%%%%%%%%%%%%%%%%%%%%
\section{Discussion and conclusions}
\label{sec:9}
%%%%%%%%%%%%%%%%%%%%%%%%%%%%%%%%%%%%%%%%%%%%%%%%%%%%%%%%%%%

From the global analysis of the dependence of Casimir energy on the type of boundary conditions 
performed throughout this paper we can extract some consequences of  
physical interest. First, we have shown the existence of new boundary conditions which are fully consistent with
fundamental requirements of quantum field theory. Some of these conditions can be experimentally implemented.
Second, the spectral approach to the calculation of Casimir energy has been revealed as a very useful tool, not only in cases where it
can be analytically implemented but also for achieving a very efficient numerical calculation in any case.
In this way we analysed the global properties of Casimir energy ${E}^{}_c(U)$ as a function in the space of
all consistent boundary conditions $\mpro_F$.

We have univocally  characterised which boundary conditions induce an attractive Casimir force and which ones a repulsive 
Casimir force.  However, we have been unable to find the underlying physical arguments that characterise the  boundary conditions that induce  attractive or repulsive  the Casimir forces, although the algorithm used in the paper provides the simplest mechanism to determine its character. In particular, we have fully characterised the 3-dimensional family of boundary conditions that
are in the interface between the attractive and repulsive regimes. This family of Casimirless boundary conditions has a very special property: that their Casimir force vanishes, which may have some interest for physical applications.

We have confirmed that all boundary conditions corresponding to identical bodies are always attractive in agreement with the Kenneth-Klich theorem. In fact,  we have shown that  the same behaviour holds for bodies with slightly different boundary conditions. A result that follows from the continuity of the Casimir Energy in the space of boundary conditions. In general the repulsive behavior requires rather different boundary conditions for the two plates.

%Comparison of
% these results with those 
%of their 1+1 dimensional analog highlights that the singularity at $\alpha=0$ in the first derivative of the Casimir energy for the $1+1$-dimensional case disappears in dimension $3+1$. 
%The disappearance of the this special type of (weak) first order phase transition also occurs in 2+1 dimensional case.

It is of note that the cusp singularity of pseudo-periodic boundary conditions in 1+1 dimensions   at $\alpha=0$ has  disappeared in 2+1 dimensions (Figure \ref{mergedpp}). This means that the (weak) first order phase transition that occurs in $1+1$ dimensions becomes a weaker higher order phase transition in 2+1 dimensions, and we do not observe any kind of phase transition in 3-dimensional
systems at $\alpha=0$.

The strong convergence properties of the spectral integral that defines the Casimir energy also implies that the 
Casimir energy function in $\mpro_F\subset U(2)$ is an holomorphic function when restricted to the interior of
the domain. However some singular points can appear at the border of such a space (see Figure \ref{mergedpp}),  because beyond
that border consistency of field theory fails. This property has some physical consequences because if 
${E}^{}_c(\alpha,\,\beta,\,{\bf n})$ is holomorphic inside $\mpro_F\cap U(2)$ its extremal values have to be
attaint at the boundary according to the minimun-maximum principle. This explains why we found 
the extremal points on the corner of the rhombus.

The extremal points correspond in 1+1, 2+1 and 3+1 dimensions to periodic (minimum) and anti-periodic (maximum)
boundary conditions. This property appears to hold in higher dimensions which motivates an interesting conjecture.
Periodic boundary conditions  always generate the strongest attractive Casimir force between the plates 
whilst anti-periodic conditions generate the strongest repulsive force. In fact, the conjecture can be proven  using 
inequalities similar to equation (\ref{ineq}).
 
On the other hand there is an interesting mismatch between the gradient flow generated by the Casimir energy function and the renormalisation group flow given by  \cite{adj2}
\begin{equation}
 \Lambda\, U_\Lambda^\dagger \partial_{\Lambda} U_\Lambda= \frac12 \left(U^\dagger_\Lambda -U_\Lambda\right)
\end{equation}
The fixed points of the RG flow correspond to conformally invariant boundary conditions. However due to the Casimir effect
these points are not completely stable. The existence of this property for periodic and anti-periodic boundary conditions 
is well know from the analysis of the
conformal anomaly in 1+1 dimensions boosted by string theory.
Only a small family of boundary conditions (\ref{fixed}) 
are conformally invariant and without Casimir force \cite{ischia}.  They can be identified as the boundary conditions sitting at
the left and right corners of the rhombus satisfying equation  (\ref{doce}). The field theories with these boundary conditions are
conformally invariant and anomaly free, i.e. the vacuum energy vanishes. 
This opens a new approach to the study of string theory in  non-critical dimensions which deserves further study.
The stability  under these boundary conditions of interacting field theories is also an interesting open question.

Finally, it will be very interesting to generalise the previous analysis to gauge  field theories and obtain the dependence  of the vacuum energy of gauge  field theories on the most general type of boundary conditions from a global perspective.

%%%%%%%%%%%%%%%%%%%%%%%%%%%%%%%%%%%%%%%%%%%%%%%%%%%%%%%%%%%
\section*{Acknowledgements}
%%%%%%%%%%%%%%%%%%%%%%%%%%%%%%%%%%%%%%%%%%%%%%%%%%%%%%%%%%%

We thank M. Bordag,  I. Cavero,  K. Kirsten, G. Marmo, D. Vassilevich and  J. Mateos Guilarte, for enlightening discussions on several aspects of the Casimir effect. J. M. M. C. would like to thank S. Ratcliffe for her english support.  This work has been supported by the Spanish DGIID-DGA grant 2009-E24/2, the Spanish  MICINN grants FPA2009-09638 and CPAN-CSD2007-00042, the German DFG grant BO 1112/18-1, and the European Union ESF Research Network CASIMIR.

\end{document}